\documentclass[12pt]{article}

\usepackage{epsfig, amssymb, amsmath}
\DeclareGraphicsExtensions{.jpg, .eps}
\newcommand{\be}{\begin{equation}}
\newcommand{\ee}{\end{equation}}
\newcommand{\bea}{\begin{eqnarray}}
\newcommand{\eea}{\end{eqnarray}}
\newcommand{\ba}{\begin{array}}
\newcommand{\ea}{\end{array}}

\def\bbox{{\,\lower0.9pt\vbox{\hrule \hbox{\vrule height 0.2 cm
\hskip 0.2 cm \vrule height 0.2 cm}\hrule}\,}}
\newcommand{\dsl}{\pa \kern-0.5em /}

\font\mybb=msbm10 at 10pt
\def\bb#1{\hbox{\mybb#1}}

\def\bE {\bb{E}}

\def\appendix#1{
  \addtocounter{section}{1}
  \setcounter{equation}{0}
  \renewcommand{\thesection}{\Alph{section}}
  \section*{Appendix \thesection\protect\indent \parbox[t]{11.15cm}
  {#1} }
  \addcontentsline{toc}{section}{Appendix \thesection\ \ \ #1}
  }

\begin{document}

\begin{flushright}
\small
DAMTP-2007-24\\
{\bf hep-th/0703276}\\
\date \\
\normalsize
\end{flushright}
\thispagestyle{empty}
%%%%%%%%%%%%%%%%%%%%%
%\vskip 1cm
%\centerline{\Large {\bf DRAFT}}
%\vskip 1cm
%%%%%%%%%%%%%%%%%%%

\begin{center}

%title

\vspace{.7cm}

{\Large {\bf Axion-Dilaton Domain Walls \\}
\smallskip
{\bf and Fake Supergravity}}

\bigskip

\vspace{1.2cm}

%authors
{\large Julian Sonner and Paul K. Townsend}

\vskip 1truecm

\small
{Department of Applied Mathematics and
Theoretical Physics\\
Centre for Mathematical Sciences, University of Cambridge\\
Wilberforce Road, Cambridge, CB3 0WA, UK}
\vspace{.7cm}

%%%%%%%%%%%%%%%%%%%%%%%%%%%%%%%%%%%%%%%%%%%%%%%%%%%%%%%%%%%%%%
%%%%%%%%

\vskip2cm
{\bf Abstract}

\end{center}

\begin{quotation}

\small

Dynamical systems methods are used to investigate domain-wall solutions of a  two-parameter family of models in which gravity is coupled to an axion, and to a dilaton with an exponential  potential of either
sign. A complete global analysis is  presented for (i) constant axion and (ii) flat walls, including a study of bifurcations and  a new exact domain-wall solution with non-constant axion.  
We reconsider `fake supergravity' issues in light of these results. We show, by example, how domain walls determine multi-valued  superpotentials that  branch at stationary points that are not stationary points of the potential, and we apply this result to potentials with anti-de Sitter vacua. We also show by example  that  `adapted'  truncation to a single-scalar model  may be inconsistent, and we propose
a `generalized' fake supergravity formalism that applies in some such cases.

\end{quotation}

\newpage
\setcounter{page}{1}
\tableofcontents
%%%%%%%%%%%%%%%%%%%%%%%%%%%%%%%%%%%%
\section{Introduction} \setcounter{equation}{0}
%%%%%%%%%%%%%%%%%%%%%%%%%%%%%%%%%%%%

In a recent paper \cite{Sonner:2005sj} we addressed the problem of  domain wall solutions of  the coupled Einstein-dilaton equations, in $d$ spacetime dimensions, using dynamical systems methods imported from studies of cosmological solutions of the same model \cite{Halliwell:1986ja,Burd:1988ss}. We recovered efficiently and simply many of the previously known exact results on dilaton domain walls and found some new ones. In addition, we obtained a qualitative overview of the entire space of domain wall solutions in this model. For example, the ``Janus'' walls  that were studied in \cite{Freedman:2003ax} were shown to have a natural interpretation as marginal bound states of  a new type of dilaton domain wall that we called a ``separatrix-wall''. This result is reminiscent of BPS solitons and it suggested a `hidden' supersymmetry of the separatrix-wall that has since been confirmed \cite{Skenderis:2006jq}. 

One purpose of this paper  is to continue our dynamical systems investigations of domain walls in a more general class  of models for which both the dilaton  field $\sigma$ and a pseudo-scalar `axion'  field $\chi$ are coupled to a $d$-dimensional metric $g$ via the Lagrangian density
\be\label{laginitial}
{\cal L} = \sqrt{-\det g}\left[ R -{1\over2}\left(\partial\sigma\right)^2 - {1\over2} e^{\mu\sigma}\left(\partial\chi\right)^2 - V\right]\, , 
\ee
where $\mu$ is an axion-dilaton coupling constant, and the potential function $V$ takes the form
\be
V= \Lambda e^{-\lambda\sigma} \, , 
\ee
where $\lambda$ is the dilaton `coupling constant',  and  $\Lambda$ a non-zero `cosmological' constant.  This is the unique form of the potential that preserves an  invariance of the equations of motion under the `dilation' $g \to e^\omega g$, for constant parameter $\omega$, by virtue of the inhomogeneous transformation  $\sigma +\omega/\lambda$ of the dilaton field. Only the sign of $\Lambda$ is relevant when $\lambda\ne0$,  as a shift of $\sigma$ is then equivalent to a scaling of $\Lambda$.  We may assume that $\lambda\ge0$ without loss of generality, as  in \cite{Sonner:2005sj},  but then $\mu$ could have either sign; its absolute value is related to the radius of curvature of the hyperbolic target space for which $(\sigma,\chi)$ are  coordinates. The $\mu=0$ case, which corresponds to a flat target space, has been investigated (in a cosmology context) in \cite{Bergshoeff:2003vb,Townsend:2004zp}.  

We have recently studied cosmological solutions for this class of axion-dilaton  model \cite{Sonner:2006yn}, following the work of  \cite{Billyard:2000cz} on a particular model  for which  flat universes were shown to expand and contract quasi-periodically in a certain non-Einstein frame. We found a range of the parameters $(\lambda,\mu)$, for which generic flat universes are eternally expanding, in Einstein frame, but undergo  a medium-time and/or late-time oscillation between acceleration and deceleration. These results were one motivation for the present work because if one allows for both signs of the potential then the family of dynamical systems governing cosmological solutions is the same as the family  of dynamical systems governing domain-wall  solutions (illustrating the general `Domain-Wall/Cosmology Correspondence'  \cite{Skenderis:2006jq}),  and  the quasi-cyclic nature of many trajectories  could imply novel behaviour for associated holographic renormalization group flows. 

The equations for either domain walls or (homogeneous and isotropic) cosmologies in a model with $n$ scalar fields define an autonomous  dynamical system of dimension $2n+1$, once account is taken of the reparametrization constraint. Reparametrization invariance implies that the physical phase space is actually only $2n$-dimensional but the dynamics on this space is not necessarily autonomous.  Thus, even for $n=1$ we should expect to have to consider a 3-dimensional system,  and  a 5-dimensional system for $n=2$.
However, a reduction to an autonomous 2-dimensional system is possible for  $n=1$  when the
potential is an exponential, as is well-known; this case arises in the present context from the restriction to constant axion. Less well-known is the fact that the 5-dimensional dynamical system for $n=2$ models of the above axion-dilaton type has a 2-dimensional autonomous subsystem  when the restriction is made to flat domain walls, or cosmologies, and the behaviour is essentially determined by this subsystem.  

Here we present a detailed study of the 2-dimensional dynamical systems corresponding to 
(i) constant axion and (ii) flat domain walls, for either sign of $\Lambda$ and as a function of the two parameters $(\lambda,\mu)$. Dynamical systems methods are particularly powerful in these cases because trajectories are easily visualized, and chaotic dynamics is excluded. Case (i) has been investigated previously in \cite{Halliwell:1986ja,Burd:1988ss,Townsend:2004zp} (in the context of cosmology) and in our previous 
paper \cite{Sonner:2005sj}  (in the context of domain walls) but here we present a complete global analysis. In particular, we show that the global phase space, allowing for both signs of $\Lambda$,  is a sphere. We also extend the analysis of  bifurcations to those that occur `at infinity', which corresponds to a great circle on the sphere separating the trajectories with $\Lambda>0$ from those with $\Lambda<0$. Case (ii) has been investigated previously  in the context of cosmology \cite{Sonner:2006yn,Billyard:2000cz} for one sign of $\Lambda$. Here we present a complete global analysis that includes both signs of $\Lambda$. One new result is an exact  flat domain-wall `scaling' solution with non-constant axion field. It corresponds to a fixed point of the dynamical system and for $\Lambda<0$ it is the domain-wall analog of the $\Lambda>0$ axion-dilaton cosmological solution found in \cite{Sonner:2006yn}. 

For some values of the parameters $(\lambda,\mu)$ and a choice of the sign of $\Lambda$, the model defined by (\ref{laginitial}) may be a truncation of a supergravity theory. For example,  the Freedman-Schwarz $N=4$, $d=4$, supergravity \cite{Freedman:1978ra} has  $2\lambda=-\mu =2$, and\footnote{This model was incorrectly identified in \cite{Sonner:2006yn} as one with $\Lambda>0$.} $\Lambda<0$, and in this context one may ask whether any given domain wall solution preserves some fraction of the supersymmetry. A necessary condition for partial preservation of supersymmetry  is that the domain wall solution admit  a Killing spinor, which is a spinor field satisfying the equation that results from the requirement of vanishing supersymmetry variation of the gravitino field(s); this equation depends not only on the spacetime metric but also on the (pseudo)scalar fields through a `superpotential', which determines the scalar potential.  More generally, one can extend the notion of a `supersymmetric' domain wall solution  beyond the supergravity context by allowing the superpotential  to be any function that yields the potential according to a $d$-dimensional generalization of the formula that applies in the supergravity case \cite{Townsend:1984iu}.  Domain wall solutions admitting  Killing spinors with respect to such a superpotential \cite{Skenderis:1999mm,DeWolfe:1999cp} are then said to be `fake supersymmetric' solutions of a `fake supergravity' \cite{Freedman:2003ax}.  The relevance of this idea is that `fake' supersymmetry suffices for classical stability.  

Another purpose of this paper is to use the axion-dilaton models as a `laboratory' for further 
investigations of  fake supersymmetry, although we will also have  more to say about the simpler  
Einstein-dilaton model of \cite {Sonner:2005sj}.  It was shown in \cite{Freedman:2003ax,Sonner:2005sj} that `almost all'  flat domain-wall solutions of any single-scalar model are fake supersymmetric because the solution can be used to construct a superpotential with respect to which the first-order Killing-spinor integrability equations are satisfied. This result was extended to  particular curved domain walls in \cite{DeWolfe:1999cp} and to `almost'  all curved domain  walls in \cite{Skenderis:2006jq}. The qualification `almost' arises from a monotonicity  condition. Typically, this condition is violated at 
isolated points (although non-isolated accumulation points occur in walls that are asymptotic to perturbatively  unstable anti-de Sitter vacua \cite{Skenderis:2006rr}). In this case the associated 
superpotential is multi-valued and the domain wall is `piecewise' supersymmetric with respect to it \cite{Skenderis:2006jq}. We illustrate this phenomenon here with a simple double-valued superpotential that can be found from an exact flat domain wall solution of  \cite{Sonner:2005sj}. This example illustrates another significant feature,  intrinsic to the phenomenon: the branch point of the superpotential is a stationary point that is not a stationary point of the potential.  We use the insights gained from this example to discuss general potentials with adS vacua, showing that there again exist conditions under which some superpotential will be double-valued and hence not defined for all values of the scalar 
field, as was recently argued for a particular potential \cite{Amsel:2007im}. 

Fake supergravity models with multiple scalar fields were  investigated briefly in \cite{DeWolfe:1999cp}
and more extensively in \cite{Celi:2004st}, where it was observed that for any given domain-wall solution there exist, at least locally,  `adapted'  target space coordinates such that all scalar fields but one are constant \cite{Celi:2004st}. The given solution is thus manifestly a solution of the single-scalar model obtained by the obvious truncation in the new variables. It is also (if it satisfies the monotonicity condition) a supersymmetric solution of the `adapted' single-scalar model  \cite{Skenderis:2006jq}.
Thus, results on fake supersymmetry for single-scalar models can be extended simply to multi-scalar models. However, there is a difficulty with this extension (in addition to possible global problems arising from the local nature of the `adapted' target space coordinates). To address stability of a given solution one needs to consider  solutions  `nearby'  the given one. It may happen, if the single-scalar truncation is inconsistent, that there are no nearby solutions of the single-scalar model that are also solutions of the multi-scalar model, in which case the fake supersymmetry of the domain wall as a solution of the single-scalar model is irrelevant to its stability.  The {\it consistency} of the single-scalar truncation is therefore important. Consistency is manifest for the restricted class of multi-scalar models considered in \cite{DeWolfe:1999cp} and it was shown in  \cite{Celi:2004st} that one-scalar truncations adapted to  {\it supersymmetric} domain-wall solutions of $d=5$ supergravity are consistent. 

In view of these facts, it would be a  rather unsatisfactory state of affairs, violating the sprit of fake supergravity, if a truncation adapted to a {\it non-supersymmetric} solution of some supergravity theory were to be consistent, so it is natural to conjecture  that  this never happens, i.e. that  {\it any one-scalar truncation adapted to any non-supersymmetric domain-wall solution of some supergravity theory is inconsistent}. For a solution that is not known explicitly we know of no way to determine whether the associated adapted truncation is consistent, so we are not in a position to prove the conjecture. 
However, the new exact domain-wall solution with non-constant axion mentioned earlier allows us to test it. The target space coordinates adapted to this solution can be found explicitly and it turns out that the associated single-scalar truncation is inconsistent.  If the range of the parameters $(\lambda,\mu)$ for which this solution exists are compatible with the consistent embedding of the model in a supergravity theory then the conjecture states that the solution must be non-supersymmetric in this context. As we shall show, there is a consistent truncation of minimal $d=4$ supergravity to the $d=4$
axion-dilaton model provided that $\lambda\mu=-2$, and this restriction on the parameters is compatible with the existence of the new exact domain-wall solution, which may therefore be viewed as a supergravity solution. We then show, in agreement with the above conjecture, that  it is {\it not} supersymmetric in this context. Moreover, it  remains non-supersymmetric as a solution of a $d$-dimensional  axion-dilaton fake supergravity that we propose as a generalization of the standard $d$-dimensional one-scalar fake supergravity formalism.

\subsection{Preliminaries}
\label{sec:prelim}

As in \cite{Sonner:2005sj}, we introduce the $d$-dependent constants
\be
\alpha = \sqrt{d-1\over 2(d-2)}\, , \qquad \beta = 1/\sqrt{2(d-1)(d-2)}\, . 
\ee
and express the spacetime metric in terms of a function $\varphi(z)$ through the domain-wall ansatz
\be\label{ansatz}
ds^2 = e^{2\alpha\varphi} f^2 dz^2 + e^{2\beta\varphi} d\Sigma_k^2\, ,
\ee
where $f(z)$ is an additional arbitrary function that allows us to maintain invariance under 
$z$-reparametrizations,  and $d\Sigma_k^2$ is the metric of a ($d-1$)-dimensional  maximally symmetric spacetime with inverse radius of curvature $k$;  i.e. such that the scalar curvature is $k(d-1)(d-2)$. We may restrict $k$ to take the values $-1,0,1$, and we then have a domain wall spacetime that is foliated by anti-de Sitter (adS), Minkowski or de Sitter (dS) spacetimes, respectively. 
Invariance of the solution under the isometries of the metric implies that  the fields $(\sigma,\chi)$ can depend only on $z$. The field equations now reduce to equations of motion, and a constraint, for the variables $(\varphi,\sigma,\chi)$ that are the Euler-Lagrange equations of the effective Lagrangian\footnote{Because of the inclusion of the function $f$ in the ansatz, this effective Lagrangian can also be obtained by substitution of the ansatz in the Lagrangian 
density (\ref{laginitial}), whereas this is true only for $k=0$ for an ansatz that fixes the $z$-reparametrization invariance.} 
\begin{equation}
L_{eff}= {1\over2} \left[ f^{-1} \left(\dot\varphi^2 -  \dot\sigma^2 - e^{\mu\sigma} \dot \chi^2 \right) + fe^{2\alpha\varphi}\left( k\beta^{-2} e^{-2\beta\varphi} -2\Lambda e^{-\lambda \sigma}\right)\right]\, .
\end{equation}
The equations of motion for the gauge choice
\be\label{gc}
f= e^{-\alpha\varphi + {1\over2}\lambda\sigma}
\ee
are equivalent to the equations
\bea\label{mastereq}
\ddot\varphi  &=& -\alpha\dot\varphi^2 + {1\over2}\lambda \dot\varphi\dot\sigma -2\alpha\Lambda + {k\over 2\alpha\beta^2} e^{\lambda\sigma -2\beta\varphi} \, ,\nonumber\\
\ddot\sigma &=& -\alpha\dot\sigma\dot\varphi +{1\over2} \left(\lambda-\mu\right)\dot\sigma^2 + {1\over2}\mu \dot\varphi^2 + \left(\mu-\lambda\right)\Lambda 
\eea
and the constraint
\be\label{mastercon}
\dot\chi^2 = e^{-\mu\sigma}\left[\dot\varphi^2 -\dot\sigma^2 + 2\Lambda - {k\over \beta^2}e^{\lambda\sigma -2\beta\varphi}\right] \, . 
\ee
The $\chi$ equation of motion follows from differentiation of the
constraint.  

Although the gauge choice (\ref{gc}) is  convenient for our particular models, it has the disadvantage  that $z$ is not (generically) an affine parameter when $\lambda\ne0$. It is related to an affine distance parameter $\tilde z$ through the differential equation 
\be\label{affinez}
d\tilde z = e^{\frac{1}{2}\lambda\sigma(z)} dz\, , 
\ee
and the spacetime metric takes a standard form in terms of $\tilde z$; for flat walls we have
\be
ds^2 = d\tilde z^2 + e^{2\beta\varphi}ds^2\left(\bE^{(1,d-2)}\right)\, , 
\ee
where (perpetrating a slight abuse of notation to avoid the introduction of new symbols for dependent variables) $\varphi$ is now to be understood as  that function of $\tilde z$ obtained by use of the inverse function $z(\tilde z)$ in $\varphi(z)$, and similarly for $\sigma$.

For the two special cases (i) $\dot\chi=0$ and (ii) $k=0$,  
the equations (\ref{mastereq}) contain a 2-dimensional autonomous dynamical
system  for the variables
\be
u=\dot\sigma\, ,\qquad v= \dot\varphi\, . 
\ee
In the first case, two values of $\lambda$, the `critical'  value $\lambda_c$ and the `hypercritical' value $\lambda_h$ are special because they correspond to bifurcation points. In our conventions 
\be\label{lambdas}
\lambda_c = 2\sqrt{\alpha\beta} = \sqrt{\frac{2}{d-2}}\, ,\qquad
\lambda_h = 2\alpha = \sqrt{d-1}\, \lambda_c \, . 
\ee

\section{Constant axion}
\label{sec:constantchi}
\setcounter{equation}{0}

For $\dot\chi=0$,  the equations (\ref{mastereq}) and constraint (\ref{mastercon}) yield the autonomous dynamical system
\bea\label{DS1}
\dot u &=& -\alpha uv +{1\over2}\lambda u^2 -\lambda\Lambda\nonumber\\
\dot v &=& -\beta v^2 -{1\over 2\alpha} u^2 + {1\over2}\lambda uv - 2\beta\Lambda
\eea
and the constraint
\be\label{constraint}
v^2-u^2+2\Lambda = {k\over\beta^2}e^{\lambda\sigma -2\beta\varphi}\, . 
\ee
The constraint shows that the $k=-1$ trajectories are separated from the $k=1$ trajectories by the $k=0$ hyperbola $v^2-u^2 + 2\Lambda=0$.  As the absolute value of $\Lambda$ is irrelevant for $\lambda>0$, we effectively have two one-parameter families of dynamical systems, one for $\Lambda>0$ and another for $\Lambda<0$.  In the absence of limit cycles, the phase-plane portraits of these two systems are determined by the fixed points.  Poincar\'e index considerations show that limit cycles cannot arise for $\Lambda>0$, nor  for $k\le 0$ \cite{Sonner:2005sj}. This observation leaves open the possibility of $k=1$ limit  cycles in the $\Lambda<0$ system, but in this case the constraint (\ref{constraint}) implies that $v^2>0$, and hence that $\varphi(z)$ is a `monotone function', which 
is incompatible with the existence of a limit cycle. 

We now present an analysis of the global structure of the $\Lambda>0$ and $\Lambda<0$ dynamical systems. As we shall show, individually the global phase space for each system is a disc, but the two discs can be viewed as two hemispheres of a topological sphere\footnote{This result was stated without proof in \cite{Sonner:2006yn}, where phase planes for the two hemispheres were sketched for some
cases of relevance to cosmic acceleration. Only $\Lambda>0$ cosmologies were considered
in \cite{Sonner:2006yn}, since cosmic acceleration is possible only in this case, but it should be remembered that this corresponds to $\Lambda<0$ in the domain wall context, and 
that the sign of $k$ is different.}

\subsection{Global structure}

Define new phase-plane coordinates $(x_+,y_+)$ by
\begin{equation}
(x_+,y_+) = \frac{1}{\sqrt{u^2 +v^2 + 2\Lambda}} \, (u,v)\, , \qquad 
(u,v) = \sqrt{\frac{2\Lambda}{1-x_+^2-y_+^2}}\  (x_+,y_+)\, . 
\end{equation}
This maps the $k=0$ hyperbola $u^2-v^2= 2\Lambda$ into the lines $2x_+^2=1$. 
For $\Lambda>0$ the entire $(u,v)$ plane is mapped to the interior of the unit circle in the 
$(x_+,y_+)$ plane. For $\Lambda<0$ the domain of the map is the exterior of the circle of radius
$\sqrt{2|\Lambda|}$ in the $(u,v)$ plane, which is mapped to the exterior of the unit circle in the $(x,y)$ plane. Thus, the new coordinates yield a description of the dynamical system that includes both
signs of $\Lambda$, with $\Lambda$ effectively changing sign on the unit circle; this is possible because the value of $\Lambda$ becomes irrelevant as the unit circle is approached. 

The equations for the dynamical system in the $(x_+,y_+)$ coordinates are 
\begin{align}\label{xyplus}
\frac{dx_+}{d z_+} &= \frac{1}{\lambda_h} x_+y_+\left(x_+^2-y_+^2\right) + \left(x^2+y^2-1\right)\left[\frac{1}{2}\lambda
- \lambda x_+^2 + \frac{1}{\lambda_h} x_+y_+\right] \nonumber\\
\frac{dy_+}{d z_+} &=  -\frac{1}{\lambda_h} x_+^2\left(x_+^2-y_+^2\right) + 
\left(x_+^2+y_+^2-1\right)\left[ \beta  -
\lambda x_+y_+ + \frac{1}{\lambda_h} x_+^2\right]\, , 
\end{align}
where $z_+$ is a new independent variable such that
\begin{equation}
\dot z_+ = \sqrt{\frac{2\Lambda}{1 - x_+^2 - y_+^2}} \, . 
\end{equation}
Observe that the unit circle in the $(x_+,y_+)$ plane is a fixed set of the system defined by the 
equations (\ref{xyplus}). Defining polar coordinates $(r,\theta)$ by $x_+=r\cos\theta$ and $y_+=r\sin\theta$, we find that
\begin{equation}
r'\big|_{r=1} =0\, , \qquad \theta'\big|_{r=1} = -\frac{1}{\lambda_h} \cos\theta \cos 2\theta\, . 
\end{equation}
This determines the flows on the unit circle, on which there are six fixed points, two at the angles for which $\cos\theta=0$ and four at the angles for which $\cos 2\theta=0$. The trajectories for $r<1$ are those of the original dynamical system restricted to $\Lambda>0$. Similarly, the trajectories for $r>1$ are those of the original dynamical system restricted to $\Lambda<0$, but the description of this system in the $(x_+,y_+)$ coordinates is incomplete because the region in the $(u,v)$ plane with $u^2+v^2\le 2|\Lambda|$ is not covered when $\Lambda<0$.

We may remedy the incompleteness of the $(x_+,y_+)$ coordinates for $\Lambda<0$ by defining a different set of new coordinates $(x_-,y_-)$, by
\begin{equation}
(x_-,y_-) = \frac{1}{\sqrt{u^2 +v^2 - 2\Lambda}} \, (u,v)\, , \qquad 
(u,v) = \sqrt{\frac{2\Lambda}{x_-^2+y_-^2-1}}\  (x_-,y_-)\, . 
\end{equation}
This maps the $k=0$ hyperbola $u^2-v^2=2\Lambda$ into the lines $2y_-^2=1$. 
The domain of this map for $\Lambda<0$ is the {\it entire} $(u,v)$ plane, which is now mapped to the 
{\it interior} of the unit circle in the $(x_-,y_-)$ plane. It is now for $\Lambda>0$ that  the domain is the region in the $(u,v)$ plane exterior to the circle of radius $\sqrt{2\Lambda}$, and this region is mapped to the exterior of the unit circle in the $(x_-,y_-)$ plane. The equations for the dynamical system in the $(x_-,y_-)$ coordinates are
\begin{align}\label{xyminus}
\frac{dx_-}{dz_-} &= \frac{1}{\lambda_h} x_-y_-\left(x_-^2-y_-^2\right) + \left(x_-^2+y_-^2-1\right)\left[-\frac{1}{2}\lambda
+ \left(\frac{\lambda_h^2-1}{\lambda_h}\right) x_-y_- \right] \nonumber\\
\frac{dy_-}{dz_-} &=  -\frac{1}{\lambda_h} x_-^2\left(x_-^2-y_-^2\right) + \left(x_-2+y_-^2-1\right)\left[ -\beta  + \beta y_-^2 + \frac{1}{\lambda_h} x_-^2\right]\, , 
\end{align}
where $z_-$ is a new independent variable such that
\begin{equation}
\dot z_- = \sqrt{\frac{2\Lambda}{x_-^2 + y_-^2 -1}} \, . 
\end{equation}
The unit circle is again an invariant set of the system, and the dynamics on this set is exactly the same as it was for the $(x_+,y_+)$ coordinates.

We now have two new sets of variables, taking values in either the `plus' plane or  the `minus' plane. The dynamics on the `plus' plane incorporates the full dynamics on the $(u,v)$ plane for $\Lambda>0$, including trajectories `at  infinity' on this plane, and also the `large radius'  dynamics on the $(u,v)$ plane for $\Lambda<0$, again including trajectories `at infinity' . Conversely, the dynamics on the `minus' plane incorporates the full dynamics for $\Lambda<0$ and the `large radius' dynamics  for $\Lambda>0$, again including trajectories `at infinity'. Taken together, the dynamics on the `plus' and `minus' planes incorporates the full dynamics for either sign of $\Lambda$, including the trajectories `at infinity', which is mapped to the unit circle. The phase space for this dynamics is topologically a sphere.
The $(x_+,y_+)$ coordinates cover a connected open set that includes the whole of one `hemisphere' while the $(x_-,y_-)$ coordinates cover a connected  open set that includes the whole of the other `hemisphere'. The overlap is an annular region containing the unit circle on either the $(x_+,y_+)$ or $(x_-,y_-)$ plane, and the two sets of coordinates on this region are related by
\begin{equation}\label{transition}
\left(x_+,y_+\right) = \frac{1}{\sqrt{2x_-^2 +2y_-^2 -1}}\ \left(x_-,y_-\right)\, , \qquad
\left(x_-,y_-\right) = \frac{1}{\sqrt{2x_+^2 +2y_+^2 -1}}\ \left(x_+,y_+\right)\, .
\end{equation}
Note that this maps the unit circle to the unit circle. 
Note too that the line $x_+=\pm1/\sqrt{2}$ is mapped to the line
$y_-=\pm 1/\sqrt{2}$, consistent with our earlier observation that the
$k=0$ trajectories consist of the lines $2x_+^2=1$ in the `plus' plane
and $2y_-^2=1$ in the `minus' plane.

\subsection{Fixed points}

Having determined the global topology, we will find it more convenient to proceed in terms of the new variables $(x,y)$ defined by 
\begin{equation}
(x,y) = \frac{1}{\sqrt{u^2 +v^2+ 2|\Lambda|}} \, (u,v)\, , \qquad 
(u,v) = \sqrt{\frac{2|\Lambda|}{1-x^2-y^2}}\ (x,y)\, . 
\end{equation}
These new variables coincide with $(x_+,y_+)$ for $\Lambda>0$ and with $(x_-,y_-)$ for $\Lambda<0$, which means that the dynamical system on the $(u,v)$ plane is mapped into the interior of the unit circle for either sign of $\Lambda$. We thus have two dynamical systems defined in the unit disc in the $(x,y)$ plane, and we distinguish between them according to the sign
\begin{equation}
\eta=\text{sgn}\Lambda\, . 
\end{equation}
The equations for these two dynamical systems are
\begin{align}\label{xyeqs}
\frac{dx}{d\zeta} &= \frac{1}{2\alpha} xy\left(x^2-y^2\right) + \left(x^2+y^2-1\right)\left[\frac{\eta\lambda}{2}
- \frac{\lambda}{2}\left(1+\eta\right)x^2 + \left(\alpha-\eta\beta\right)xy\right] \\
\frac{dy}{d\zeta} &=  \frac{1}{2\alpha} x^2\left(y^2 -x^2\right) + \left(x^2+y^2-1\right)\left[ \beta\eta -
\frac{\lambda}{2}\left(1+\eta\right)xy + \beta \left(1-\eta\right)y^2 + \frac{1}{2\alpha} x^2\right]\, , 
\nonumber
\end{align}
where $\zeta$ is a  new independent variable such that
\be
\dot \zeta = \sqrt{\frac{2|\Lambda|}{1-x^2-y^2}}\, . 
\ee

\begin{figure}[t]
\begin{center}
 \epsfig{file=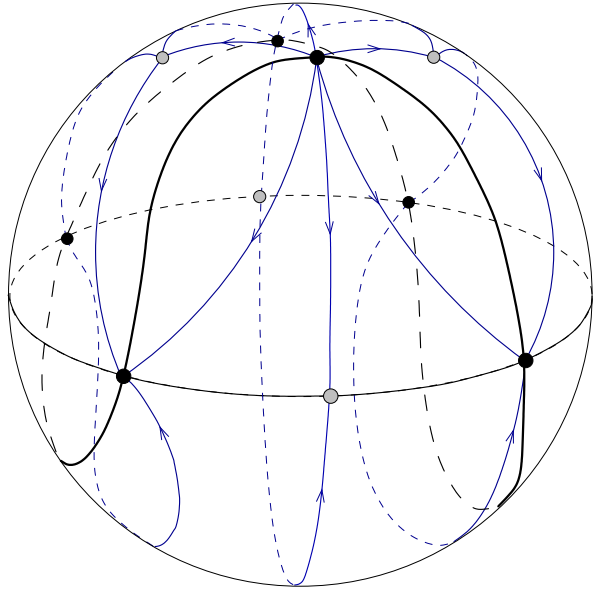, scale=0.45}
\caption{{\footnotesize The global picture is that of a topological sphere. The equator is formed by the circle at infinity and the $k=0$ invariant set forms a continuous line akin to the seam of a tennis ball. Here we show the situation for $\lambda=0$. There are a total of 10 fixed points: six nodes (black dots) and four saddles (grey dots).}\label{fig:global_sphere}}
\end{center}
\end{figure}

The fixed points on the unit circle are clearly the same for both systems. As already observed, there are six such fixed points, which we consider in pairs, in each case giving both positions and the eigenvalues of the Jacobian matrices. Firstly, we have
\begin{itemize}
\item Two saddles
\begin{equation} 
(0,1) \,:\,\left\{ 2\beta \,,\,-\frac{1}{2\alpha}   \right\} \, , \qquad 
(0,-1) \,:\,\left\{ \frac{1}{2\alpha}\,,\,-2\beta \right\}\, . 
\end{equation}
\end{itemize}
Secondly, there are four $k=0$ fixed points `at infinity':
\begin{itemize}
\item Two nodes:
\be
\pm \left(\frac{1}{\sqrt{2}},-\frac{1}{\sqrt{2}}\right) \,:\,\mp \sqrt{2}\left\{  \frac{1}{2\alpha} \, ,\, \frac{\lambda_h + \lambda}{2}   \right\}\, . 
\ee
\item Two that are nodes  for $\lambda < \lambda_h$ and saddles for $\lambda>\lambda_h$:
\be\label{saddlenodes}
\pm \left(\frac{1}{\sqrt{2}},\frac{1}{\sqrt{2}}\right) \,:\,\pm \sqrt{2}\left\{  \frac{1}{2\alpha} \, ,\, \frac{\lambda_h - \lambda}{2}   \right\} \, . 
\ee
These two fixed points are non-hyperbolic when $\lambda=\lambda_h$. 
\end{itemize}
From the global perspective, the last four fixed points are intersections of the great circle `at infinity'  on  the  spherical phase space with the continuous curve on the sphere defined by the combined $k=0$ trajectories of the $\Lambda>0$ and $\Lambda<0$ systems. If the sphere is viewed as the surface of a tennis ball then  this curve can be viewed as its seam. This is illustrated for $\lambda=0$ in Fig. \ref{fig:global_sphere}

The structure of trajectories inside the unit circle depends on the sign of $\Lambda$ so we now consider these two cases separately:

\subsubsection{$\Lambda <0$}

Setting $\eta=-1$ in (\ref{xyeqs}) we have
\begin{align}
\frac{dx}{d\zeta} &= \frac{1}{\lambda_h} \left(1-2y^2\right)xy + \frac{1}{2}\left(x^2+y^2-1\right)\left(2\lambda_h xy -\lambda\right)\,, \nonumber\\
\frac{dy}{d\zeta}&= \frac{1}{\lambda_h}  \left(2y^2-1\right)\left[x^2 + \beta\lambda_h\left(x^2+y^2-1\right)\right]\,.
\end{align}
There are two types of `interior' fixed point (i.e. inside the unit circle):
\begin{itemize}
\item Type (i), or $k=0$,  fixed points at 
\be
(x,y)=\pm \frac{1}{\sqrt{2}\lambda_h} \left(\lambda,\lambda_h\right)\, .
\ee
These lie within the unit circle iff  $\lambda<\lambda_h$. The eigenvalues of the Jacobian matrix
at these fixed points are
\be
\pm \frac{1}{2\sqrt{2}\lambda_h}\left\{\left(\lambda^2-\lambda_h^2\right), 2\left(\lambda^2-\lambda_c^2\right)\right\}\,.
\ee
This shows that these fixed points  become non-hyperbolic at $\lambda=\lambda_c$ (where they coincide with the `type (ii)' fixed points to be discussed below) and at $\lambda=\lambda_h$ (where they coincide with fixed points `at infinity'). 

\item Type (ii)  fixed points at
\be\label{type2}
(x,y)=\pm \frac{1}{\sqrt{\lambda^2+\lambda_c^2}}\left(2\beta,\lambda\right)\, . 
\ee
The eigenvalues of the Jacobian matrix at these fixed points are
\begin{equation}
\pm \frac{1}{2\lambda_h\sqrt{\lambda^2+\lambda_c^2}}\left\{ -\lambda + \frac{4}{(d-2)\bar\lambda}\sqrt{\bar\lambda^2 - \lambda^2}, -\lambda - \frac{4}{(d-2)\bar\lambda}\sqrt{\bar\lambda^2 -\lambda^2}\right\}\,,
\end{equation}
where 
\begin{equation}
\bar\lambda= \frac{4}{\sqrt{(d-2)(10-d)}}\,.
\end{equation}
These fixed points are saddles for $\lambda<\lambda_c$, nodes for $\bar\lambda\ge\lambda> \lambda_c$ and foci for $\lambda>\bar\lambda$. When $\lambda=\lambda_c$ they coincide with the type (i) fixed points. 

\end{itemize}

\subsubsection{$\Lambda>0$}

Setting $\eta=1$ in (\ref{xyeqs}) we have
\begin{align}
\frac{dx}{d\zeta} &=\frac{1}{2\lambda_h}\left(2x^2-1\right) \left[2xy - \lambda\lambda_h\left(x^2+y^2-1\right)\right]\, ,
\nonumber\\
\frac{dy}{d\zeta} &= \frac{1}{\lambda_h} x^2\left(2y^2-1\right) +\left(x^2+y^2-1\right)\left(\beta-\lambda xy\right)\, .
\end{align}
We may again classify fixed points into two types:
\begin{itemize}
\item Type (i), or $k=0$, fixed points at
\be
(x,y)= \pm \frac{1}{\sqrt{2}\lambda}\left(\lambda,\lambda_h\right)\, .
\ee
These fixed points are inside the unit circle if $\lambda>\lambda_h$. Their eigenvalues are the same as those of the $k=0$ fixed points for $\Lambda<0$. Since $\lambda_h>\lambda_c$, these fixed points are always hyperbolic, except when $\lambda=\lambda_h$, in which case they coincide with fixed points `at infinity'. 

\item Type (ii) fixed points lie on the line $2\beta y=\lambda x$, which is
  mapped into itself by the map (\ref{transition}) that takes the
  exterior  of the unit circle into the interior of the unit
  circle. The positions of any such fixed points are therefore given
  again by (\ref{type2}), as one may verify, but these fixed points
  are now outside the unit circle.

\end{itemize}

\subsection{Phase portraits}

We now present a selection of global phase portraits for representative values of the parameters. We show the two hemispheres separately, but they fit together to form a topological sphere in the way described earlier. Saddle points are indicated by grey dots and nodes/foci by black dots.

\newpage
\begin{figure}[h!]
  \hfill
  \begin{minipage}[t]{.45\textwidth}
    \begin{center}  
      \epsfig{file=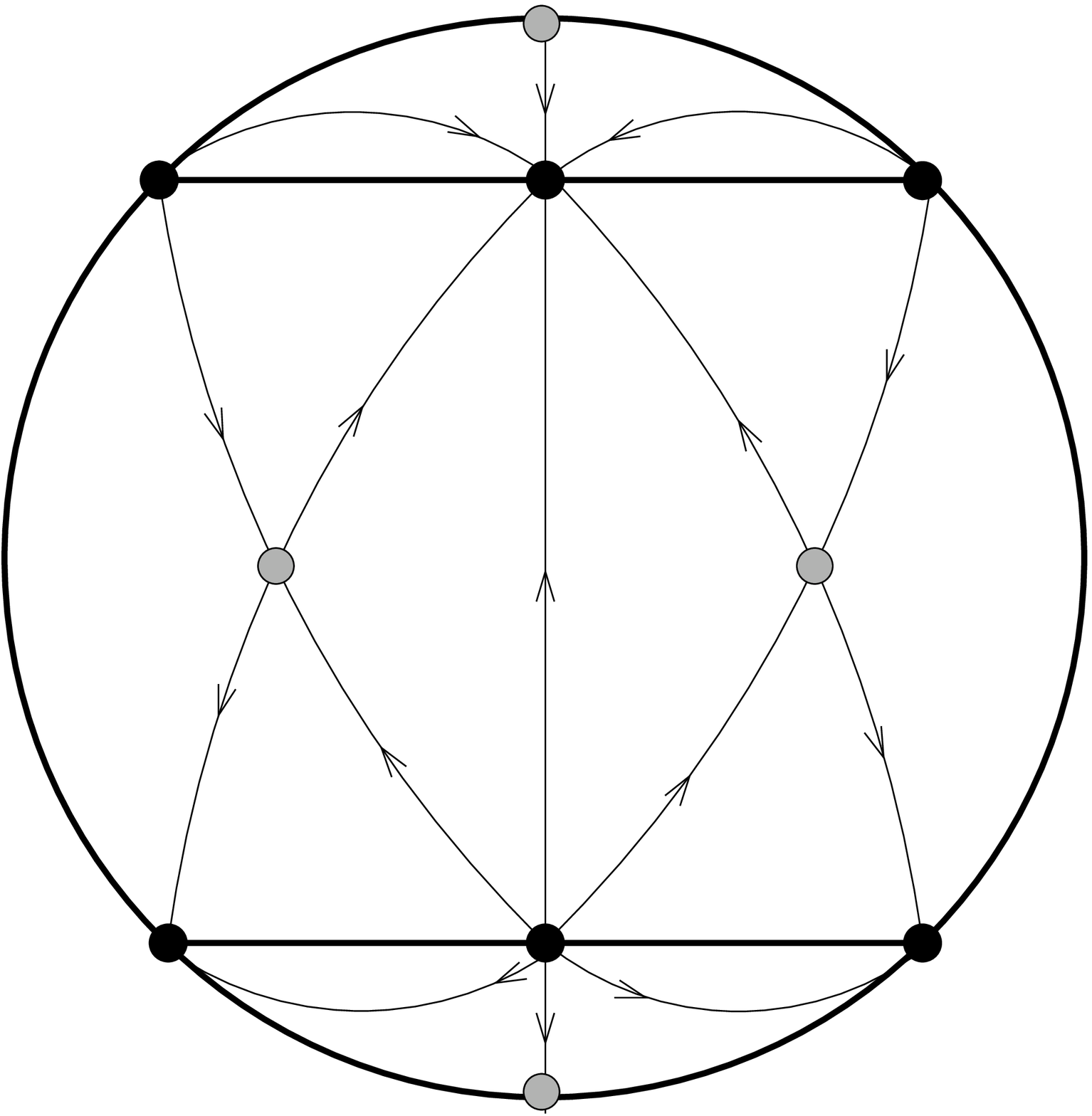, scale=0.33}
      \caption{{\footnotesize $\lambda =0 $. Hemisphere for $\Lambda<0$.}\label{fig:global1}}
    \end{center}
  \end{minipage}
  \hfill
  \begin{minipage}[t]{.45\textwidth}
    \begin{center}  
      \epsfig{file=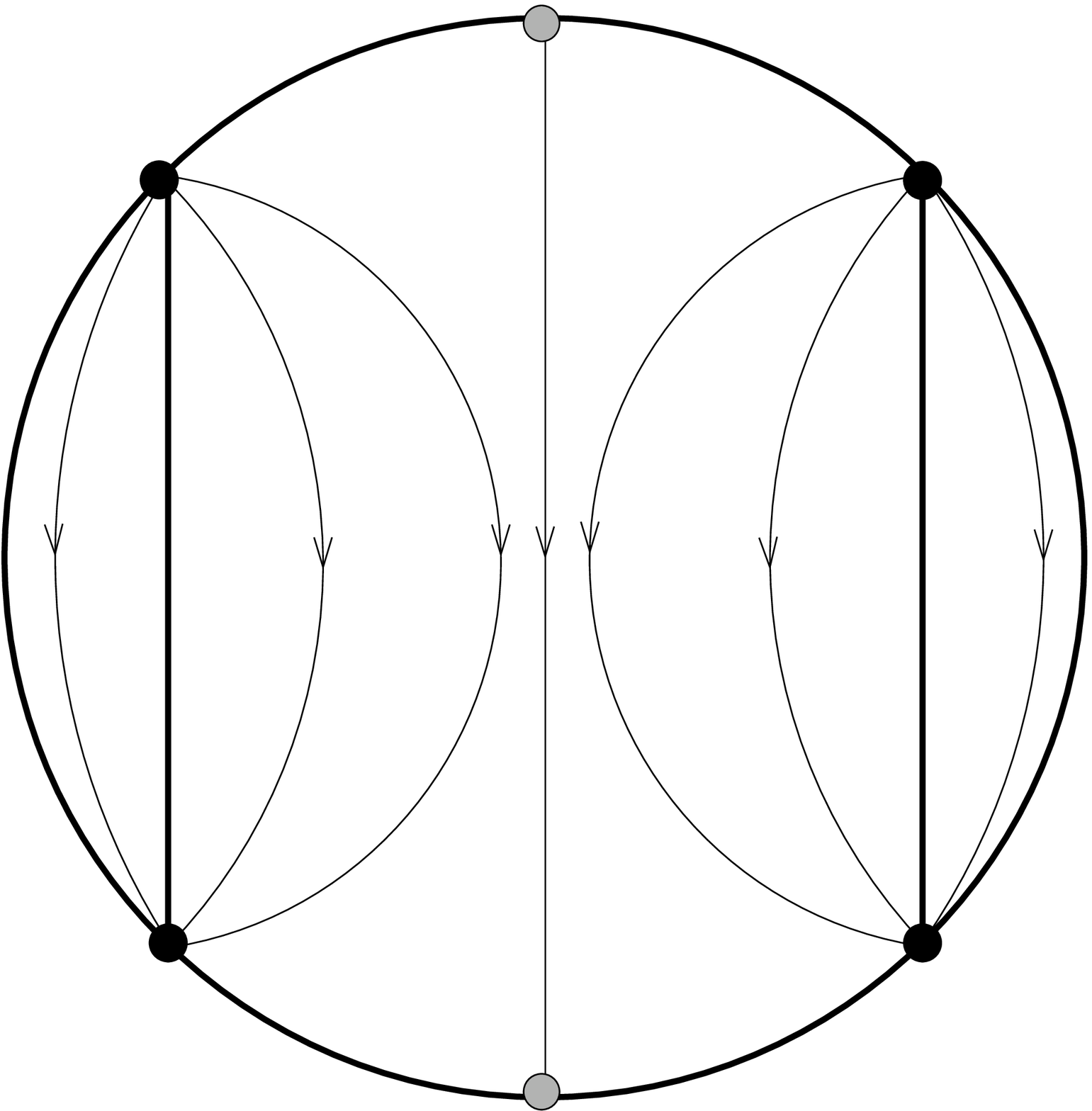, scale=0.33}
      \caption{{\footnotesize $\lambda =0 $. Hemisphere for $\Lambda>0$.}}
    \end{center}
  \end{minipage}
  \hfill
\end{figure}
\vfill
\begin{figure}[h!]
  \hfill
  \begin{minipage}[t]{.45\textwidth}
    \begin{center}  
      \epsfig{file=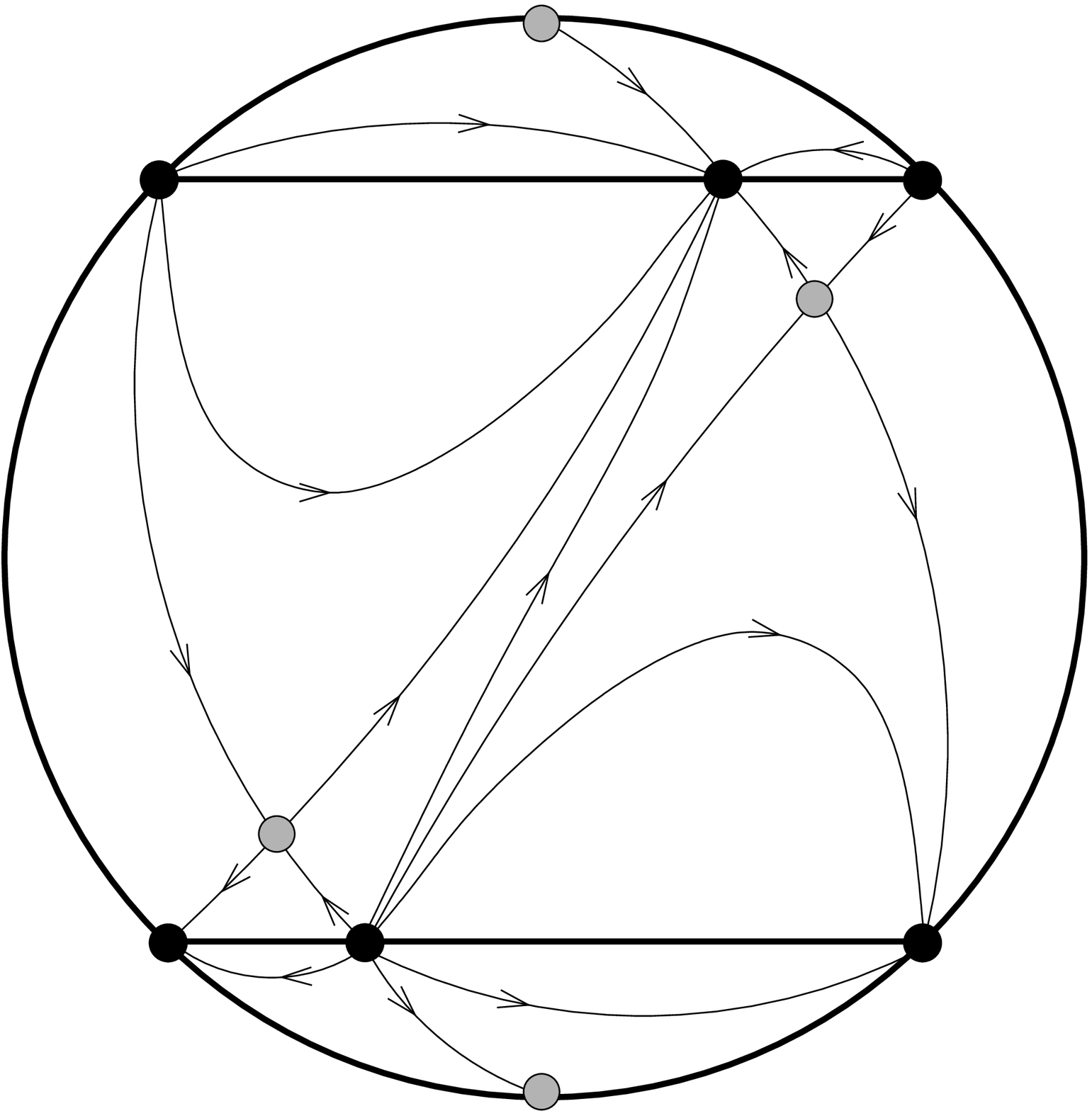, scale=0.33}
      \caption{{\footnotesize $\lambda < \lambda_c $. Hemisphere for $\Lambda<0$.}}
    \end{center}
  \end{minipage}
  \hfill
  \begin{minipage}[t]{.45\textwidth}
    \begin{center}  
      \epsfig{file=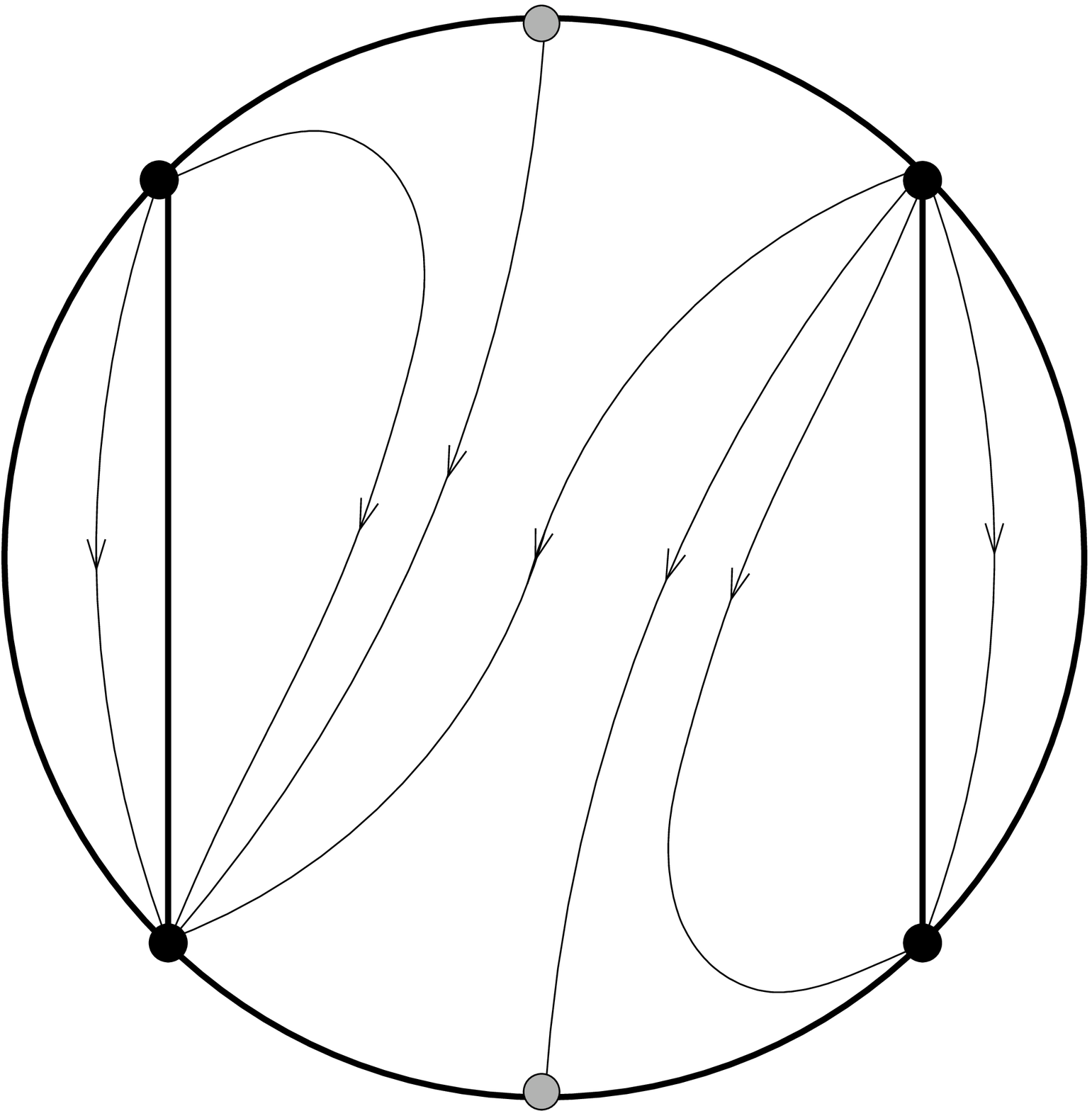, scale=0.33}
      \caption{{\footnotesize $\lambda < \lambda_c $. Hemisphere for $\Lambda>0$.}}
    \end{center}
  \end{minipage}
  \hfill
\end{figure}
\vfill
\newpage

\begin{figure}[h!]
  \hfill
  \begin{minipage}[t]{.45\textwidth}
    \begin{center}  
      \epsfig{file=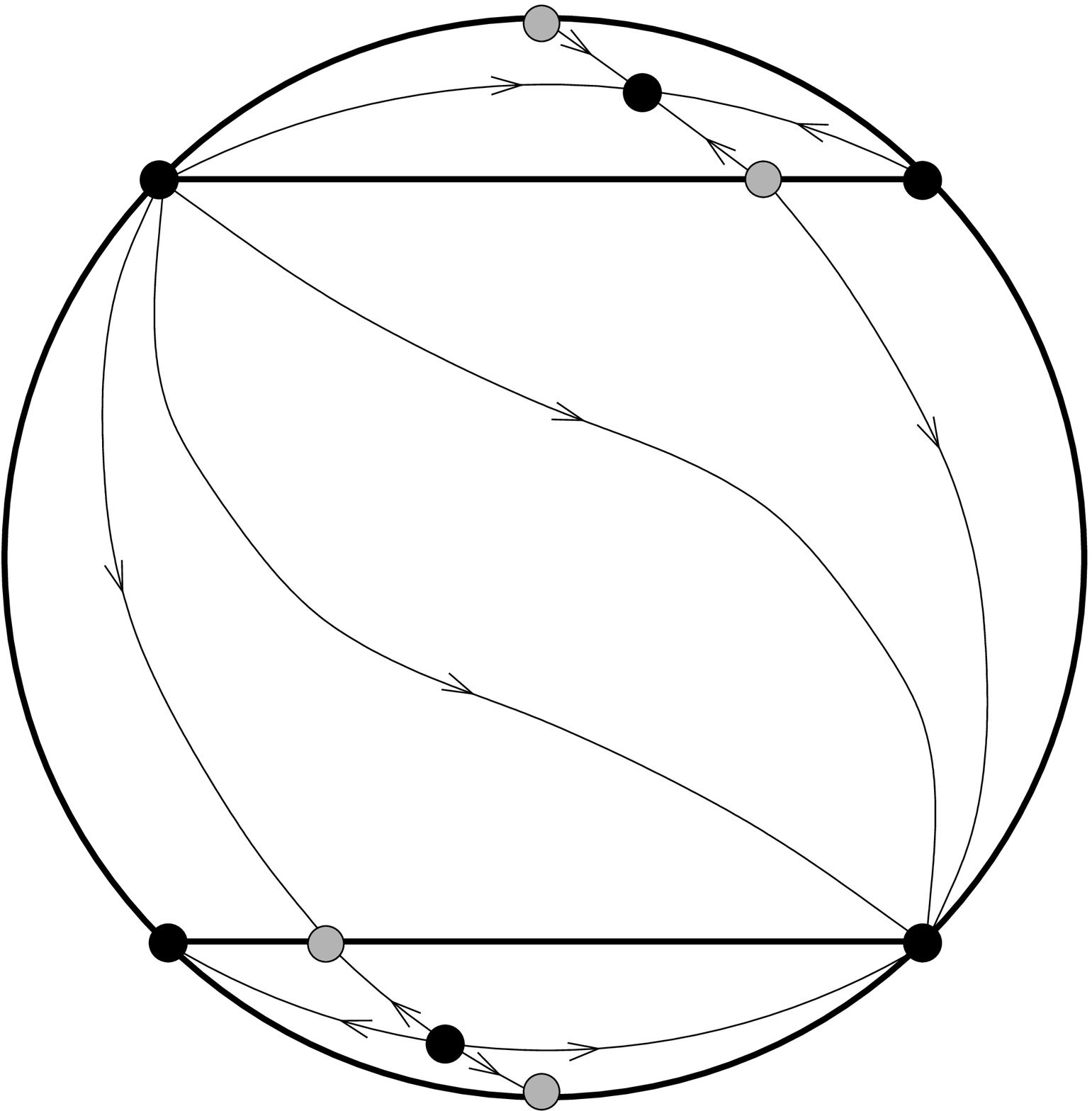, scale=0.33}
      \caption{{\footnotesize $\lambda_h>\lambda > \lambda_c $. Hemisphere for $\Lambda<0$.}}
    \end{center}
  \end{minipage}
  \hfill
  \begin{minipage}[t]{.45\textwidth}
    \begin{center}  
      \epsfig{file=fig4new.ps, scale=0.33}
      \caption{{\footnotesize $\lambda_h>\lambda > \lambda_c $. Hemisphere for $\Lambda>0$.}}
    \end{center}
  \end{minipage}
  \hfill
\end{figure}
\vfill
\begin{figure}[h!]
  \hfill
  \begin{minipage}[t]{.45\textwidth}
    \begin{center}  
      \epsfig{file=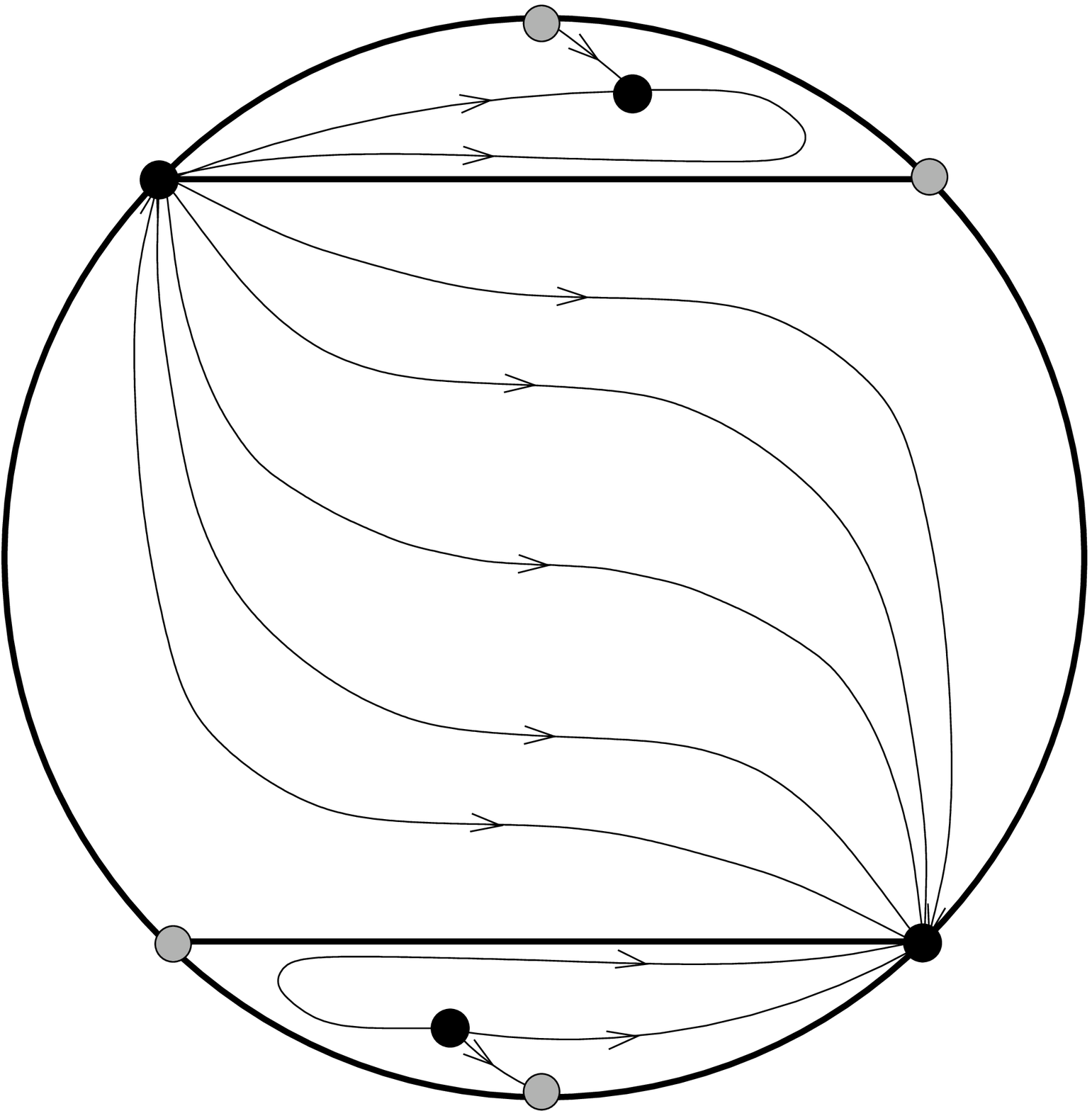, scale=0.33}
      \caption{{\footnotesize $\lambda > \lambda_h>\lambda_c $. Hemisphere for $\Lambda<0$.}}
    \end{center}
  \end{minipage}
  \hfill
  \begin{minipage}[t]{.45\textwidth}
    \begin{center}  
      \epsfig{file=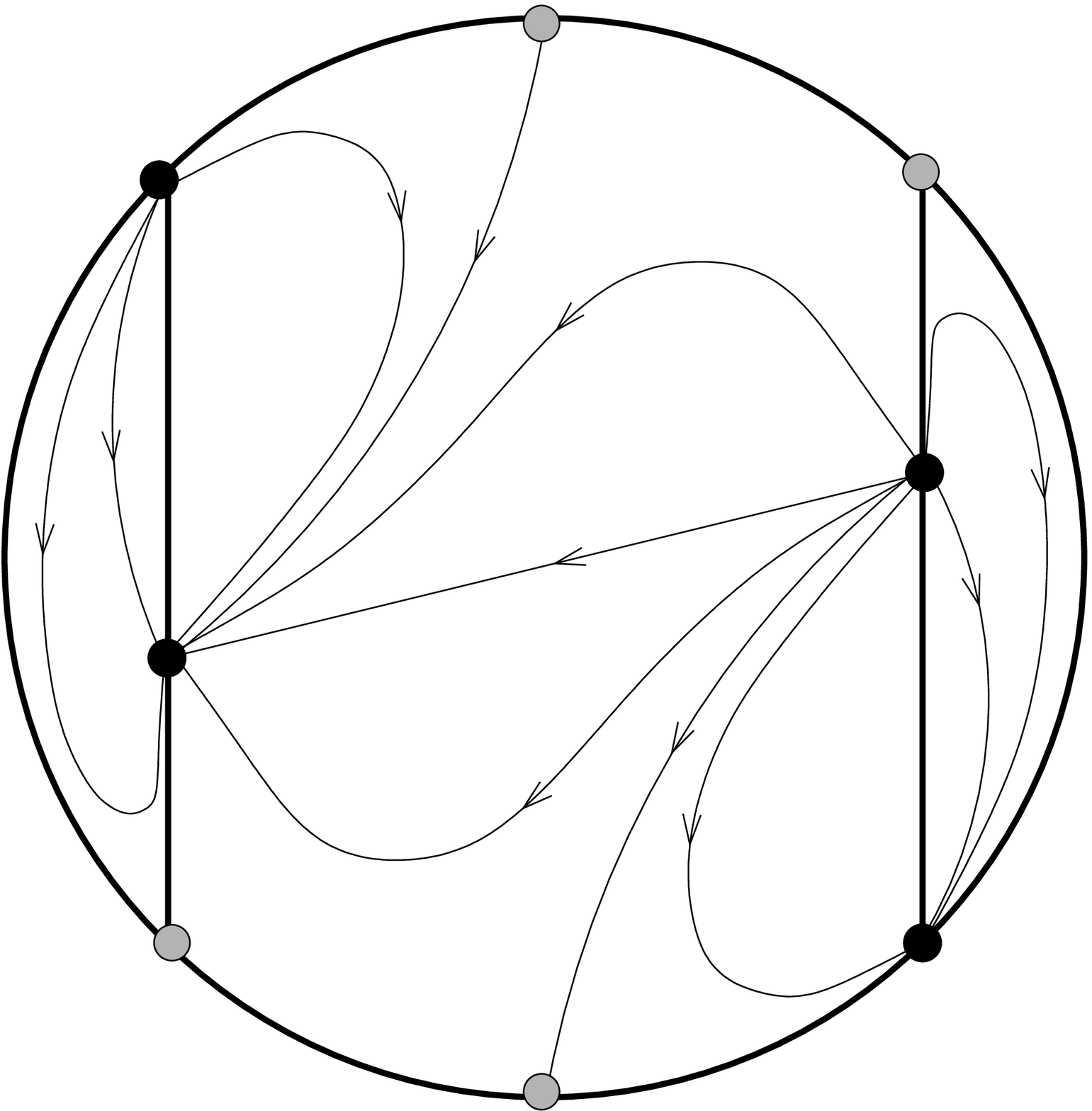, scale=0.33}
      \caption{{\footnotesize $\lambda > \lambda_h>\lambda_c $. Hemisphere for $\Lambda>0$.}}
    \end{center}
  \end{minipage}
  \hfill
\end{figure}
\vfill
\newpage

\subsubsection{Trajectories vs solutions} 
\label{sec:traj}

For either sign of $\Lambda$ the equations (\ref{DS1}) are invariant under reflection through the origin
\be
(u,v) \rightarrow -(u,v)\, . 
\ee
It follows that this transformation takes one trajectory into another trajectory, except for the trajectory through the origin, which is taken into itself. In particular, this explains why fixed points occur in pairs, since the origin is never a fixed point for non-zero $\Lambda$. This transformation is induced by the 
transformation $z\to -z$, so paired trajectories correspond to domain wall solutions that are related by this diffeomorphism, and $z\to -z$ must be an isometry of the solution corresponding to the trajectory through the origin.

These observations are nicely illustrated by the $\lambda=0$ model with $\Lambda<0$, for which all trajectories (in contrast to  domain-wall solutions) are known exactly \cite{Sonner:2005sj}. One finds that the two type (i) fixed points yield the metrics
\be
ds^2_\pm = dz^2 + e^{\pm 2\beta\sqrt{2|\Lambda|}\,  z} \, ds^2\left(\bE^{(1,d-2)}\right)\, . 
\ee
These are both locally adS metrics that are foliated by $(d-1)$-dimensional Minkowski spaces. This foliation separates the entire adS spacetime into two regions\footnote{By `adS' we mean adS and not its covering space.}, separated by a Killing horizon, and each fixed point yields the metric on one of the two regions. The trajectory through the origin yields a solution with metric
\be
ds^2= dz^2 + \frac{1}{2\beta^2|\Lambda|}\cosh\left(\beta\sqrt{2|\Lambda|}\,  z\right) 
ds^2\left(adS_{d-1}\right)\, . 
\ee
This is the same adS spacetime, but now foliated by $(d-1)$-dimensional anti-de Sitter spaces. Note the
$z\to -z$ isometry.  Given global coordinates for the $adS_{d-1}$ leaves of the foliation, the entire adS spacetime is now covered, including both regions covered by the fixed-point solutions. There are also two $k=1$ trajectories that yield a metric that is locally adS, and these correspond to two regions of adS foliated by $(d-1)$-dimensional de-Sitter (dS) spaces. These adS examples illustrates the further point that a given 
domain wall spacetime may be represented by more than one trajectory, if one regards as equivalent 
solutions that are locally diffeomorphic. 

Another distinction between trajectories and solutions emerges when we consider $\lambda>0$. It is particularly instructive to consider the models with
\be
\lambda= \lambda_n\equiv \lambda_c \sqrt{\frac{n}{d+n-2}} 
\ee
where $n$ is a non-negative integer; note that $\lambda=0$ for $n=0$ and $\lambda_n<\lambda_c$.  For $n\ge1$, solutions of the Einstein-dilaton model lift to solutions of the $(d+n)$-dimensional Einstein equations with  cosmological constant $\Lambda$. Although the higher-dimensional solution will not necessarily have a domain-wall interpretation, the $k=0$ fixed point solutions lift to the two Minkowski-foliated locally $adS_{d+n}$ metrics that are separated by a Killing horizon. These metrics do not cover the Killing horizon, which becomes a curvature singularity of the $d$-dimensional metric. However, the trajectory through the origin, which interpolates between the fixed points, yields a solution that lifts to a global covering of the $adS_{d+n}$ spacetime, so this metric has  a singularity at $z=0$, so that the $z\to -z$ transformation now interchanges two singular metrics.  It is natural to conjecture that the same result holds for all $\lambda<\lambda_c$, since the structure of trajectories is topologically the same as for $\lambda=0$. Because of the bifurcation at $\lambda=\lambda_c$, there is a topology change for $\lambda>\lambda_c$ such that the trajectory through the origin  now interpolates between $k=0$ fixed points `at infinity'. 

Another instructive example is provided by the $\lambda=\lambda_h$ model, for which there are just 
two  $k=0$  solutions, for either sign of $\Lambda$, given by\footnote{This solution was given in  \cite{Sonner:2005sj} but only for $z>0$.} 
\be
e^{\lambda_h \varphi} = |z| \exp\left(-\frac{1}{4}\Lambda\lambda_h^2 z^2\right)\, , \qquad
e^{\lambda_h\sigma} = |z|^{-1} \exp\left(-\frac{1}{4}\Lambda\lambda_h^2 z^2\right)\, . 
\ee
Consider the $\Lambda<0$ case: then $\dot\varphi>0$ for $z>0$ and $\dot\varphi<0$ for $z<0$. These two possibilities correspond to the two $k=0$ trajectories interchanged by the transformation $z\to -z$.
Let us choose the $\dot\varphi>0$ trajectory with $z>0$, and set $\Lambda=-2/\lambda_h^2$ for simplicity. Then
\be
e^{\lambda_h \varphi} = ze^{\frac{1}{2} z^2}\, , \qquad
e^{\lambda_h\sigma} = z^{-1} e^{\frac{1}{2} z^2}\, \qquad \left(\Lambda= -2/\lambda_h^2\right)
\ee
and hence
\be\label{sigzed}
\lambda_h\dot\sigma=z^{-1}\left(z^2-1\right) \, . 
\ee
We see that $\dot\sigma=0$ at $z=1$, so that the range of the function $\sigma(z)$ is restricted to
\be
\sigma(z)\ge \sigma_{min} \equiv \sigma(1) = \frac{1}{2\lambda_h}\, . 
\ee

The relation of $z$ to the affine parameter $\tilde z$ is found by integration of
\be
d\tilde z = \frac{e^{\frac{1}{4}z^2}}{\sqrt{z}} \, dz
 \ee
We see that $\tilde z\sim 2\sqrt{z}$ near $z=0$. Thus the singularity of the metric and scalar field at  $z=0$ is at finite affine distance. The domain wall solution in this case is singular. Nevertheless, it will 
be useful later.

\subsection{Bifurcations at infinity}

Bifurcations of families of dynamical systems occur when fixed points coincide for some values of the parameters, leading to a non-hyperbolic fixed point at the bifurcation point in parameter space.  There is a bifurcation in the $\Lambda<0$ family of dynamical systems when $\lambda=\lambda_c$ because in this case the type (i) and type (ii) fixed points coincide. It was shown in  \cite{Sonner:2005sj} that this is a {\it transcritical} bifurcation, in which the stability properties of two fixed points are exchanged as they
pass through each other. 

Inspection of (\ref{saddlenodes})  shows that these fixed points at infinity are non-hyperbolic when $\lambda=\lambda_h$; this is  because they then coincide with the type (i) fixed points, which move to infinity as 
$\lambda$ approaches $\lambda_h$. We now show  that this leads to another transcritical bifurcation. 
We first define new variables $(\check x,\check y)$ by
\be
x= \pm \frac{1}{\sqrt{2}} + \check x - \check y\, , \qquad
y= \pm \frac{1}{\sqrt{2}} + \check y
\ee
and we define the new parameter 
\be
s=\pm  \frac{1}{\sqrt{2}} \left(\lambda-\lambda_h\right)\, . 
\ee
We now find, for $\eta=-1$, that the system is described by the equations\footnote{The equations for $\eta=1$ are slightly different but  diffeomorphic and hence  yield equivalent results.}
\begin{align}
\frac{d\check x}{d\zeta} &= s \check x + \lambda_h \check x^2 + \frac{2}{\lambda_h} \check y\left[\left(1+2\beta\lambda_h\right)\check x -2\check y\right] + C\left(\check x,\check y,s\right)
\nonumber\\
\frac{d\check y}{d\zeta}&= \frac{\sqrt{2}}{\lambda_h} \check y + \check y L\left(\check x,\check y\right)\, , 
\end{align}
where $C$ is a polynomial function at least cubic in its arguments, and $L$ is a polynomial function that is at least linear  in its arguments. 

 There is a fixed point at the origin of the $(\check x,\check y)$ plane that becomes non-hyperbolic at $s=0$. One now considers the `extended' system in which $s$ becomes a variable subject to the trivial equation $ds/d\zeta=0$. This system has a fixed point at the origin and we are interested in the dynamics on the centre manifold through this fixed point.  Because of the structure of the above equations, the equation for this `extended' centre  manifold is simply $y=0$ through quadratic order, so the equation for the dynamics  on the extended centre manifold to quadratic order is 
\be
\frac{d\check x}{d\zeta} = s \check x + \lambda_h \check x^2\, . 
\ee
After flipping the sign of the independent variable, this equation becomes a standard one for a transcritical bifurcation.

\section{Flat walls}
\setcounter{equation}{0}

For $k=0$ we have the dynamical system
\bea\label{DS2}
\dot u &=& -\alpha uv +{1\over2}\left(\lambda-\mu\right)(u^2-2\Lambda) + 
{1\over2}\mu v^2  \, , \nonumber\\
\dot v &=& -\alpha v^2 + {1\over2}\lambda uv -2\alpha \Lambda
\eea
The constraint is
\be\label{flatconstraint}
\dot\chi^2 = e^{-\mu\sigma}\left[ v^2-u^2 +2\Lambda\right]\, , 
\ee
and it implies that we must restrict the phase space by the inequality
\be\label{ineq}
v^2-u^2 +2\Lambda \ge0\, . 
\ee
Note that when $v^2 -u^2 +2\Lambda =0$, the system reduces to the
$k=0$ case of the purely dilaton model, so the $k=0$ trajectories of 
that model are the $\dot\chi=0$ trajectories of the axion-dilaton 
model. 

We  record here that the spacetime metric for the gauge 
choice (\ref{gc}) used to obtain these equations is
\be\label{spacetime}
ds^2 = e^{\lambda\sigma} dz^2 + e^{2\beta\varphi} 
ds^2\left(\bE^{(1,d-2)}\right)\, . 
\ee

\subsection{Global Structure and Fixed Points}

Define the new coordinates
\begin{equation}
(X,Y) = \frac{1}{\sqrt{2v^2+u^2+2\Lambda}}\ (u,v)\, , \qquad
(u,v) = \sqrt{\frac{2\Lambda}{1-X^2-2Y^2}}\ (X,Y)
\end{equation}
The physical region $v^2-u^2+2\Lambda\ge0$ is mapped  onto the the interior of an ellipse:
\begin{equation}
2X^2+Y^2 \le1\, . 
\end{equation}
This ellipse can be identified with the $k=0$  `tennis ball seam' of the global phase space for constant axion  trajectories described in the previous section. 

The trajectories `at infinity' in the $(u,v)$ plane are mapped to the intersection with the physical region of the orthogonal ellipse
\begin{equation}
X^2+2Y^2=1\, .
\end{equation}
The physical region contains (i) the region with $X^2+2Y^2<1$, which is the phase space for $\Lambda>0$, and (ii) two disjoint regions with $X^2+2Y^2>1$, which collectively form the phase space for $\Lambda<0$. The equations in the new variables are
\begin{align}\label{eq:XYsystem}
\frac{dX}{d\zeta} &= \frac{1}{2}\mu\left(1-X^2\right)\left(1-2X^2-Y^2\right) + \frac{1}{2}\left(1-X^2-2Y^2\right)\left[\lambda_h XY -\lambda\left(1-2X^2\right)\right]\nonumber \\
\frac{dY}{d\zeta} & =-\frac{1}{2} \mu XY\left(1-2X^2-Y^2\right) + \frac{1}{2}\left(1-X^2-2Y^2\right)\left[2\lambda XY- \lambda_h\left(1-Y^2\right)\right]
\end{align}
where $\zeta$ is a new independent variable such that
\begin{equation}
\dot{\zeta} = \sqrt{\frac{2\Lambda}{1-X^2-2Y^2}}\, .
\end{equation}
Observe that the boundary of the physical region, $X^2+2Y^2=1$ is an invariant set, and also that the
two segments of the ellipse $2X^2+Y^2=1$ within the physical region are invariant sets.

\subsubsection{Boundary fixed points}

There are two types of fixed point on the $2X^2+Y^2=1$ boundary

\begin{itemize}
\item  
Fixed points `at infinity':  there are two fixed points on the ellipse, at $(X,Y)=(\pm1,0)$, but these are outside the physical region. The only fixed points `at infinity'  that are inside the physical region are also 
on the boundary of the physical region. There are four of them, and their positions and eigenvalues are as follows:
\begin{align}
\pm \frac{1}{\sqrt{3}}\left(1,1\right)\, : & \qquad \left\{-\frac{1}{\sqrt{3}}\mu\, , 
\ -\frac{1}{\sqrt{3}}\left(\lambda-\lambda_h\right)\right\} \nonumber\\
\pm \frac{1}{\sqrt{3}}\left(1,-1\right)\, : & \qquad \left\{-\frac{1}{\sqrt{3}}\mu\, , \ -\frac{1}{\sqrt{3}}\left(\lambda+\lambda_h\right)\right\} 
\end{align}

\item  Apart from the boundary  fixed points  `at infinity',  there is
  another pair of  fixed points on the boundary provided that
  $\lambda\ne\lambda_h$. These lie on the intersection of the boundary
  with the line $\lambda_h X=\lambda Y$. They have positions and eigenvalues
\begin{equation}\label{secondpair}
\pm\frac{1}{\sqrt{2\lambda^2+\lambda_h^2}}\ \left(\lambda,\lambda_h\right)\, :\ 
\left\{\frac{\lambda^2-\lambda_h^2}{2\sqrt{2\lambda^2+\lambda_h^2}}\, , \ \frac{\lambda^2-\lambda_h^2 -\mu\lambda}{\sqrt{2\lambda^2+\lambda_h^2}}\right\}
\end{equation}
If $\lambda<\lambda_h$ then these fixed points are on the boundary of  the $\Lambda>0$ region. If $\lambda>\lambda_h$ then there is one on each boundary of the two disjoint $\Lambda<0$ regions. 
At $\lambda=\lambda_h$ these fixed points coincide with boundary fixed points `at infinity'. 

\end{itemize}

\subsubsection{Interior fixed points}

There are a pair of fixed points with positions
\begin{equation}
\left(X,Y\right) = \pm \Delta^{-1} \left(\lambda_h,\lambda-\mu\right)\, . 
\end{equation}
where
\be
\Delta = \sqrt{\left(\lambda-\mu\right)^2 + \lambda\left(\lambda-\mu\right) + \lambda_h^2}\, . 
\ee
These are in the allowed region  $2X^2+Y^2\le1$ 
provided that\
\begin{equation}\label{interiorconstraint}
\lambda\left(\lambda-\mu\right) \ge \lambda_h^2\, . 
\end{equation}
Note that this condition requires $\lambda\ge\mu$, and that it is equivalent, for non-zero $\lambda$, to
\begin{equation}
\mu\le \mu_c \equiv \frac{\lambda^2-\lambda_h^2}{\lambda}\, .  
\end{equation}
These fixed points coincide with fixed points on the boundary when the inequality is saturated.
When it is otherwise satisfied, there are two `interior'  fixed points,  in the  $\Lambda>0$ region when $\mu>0$, and one in each of the two $\Lambda<0$ regions when $\mu<0$. 

The eigenvalues $E_\pm$ of the Jacobian matrix at an interior fixed
point are
\begin{align}\label{Evalues}
E_1 =& \pm\frac{1}{4\Delta}\left\{\mu\lambda_h +
\sqrt{8\mu\lambda\left(\lambda-\mu\right)^2
  -\mu\left(8\lambda-9\mu\right)\lambda_h^2}\right\} \, , \nonumber\\
E_2 =& \pm\frac{1}{4\Delta}\left\{\mu\lambda_h -
\sqrt{8\mu\lambda\left(\lambda-\mu\right)^2
  -\mu\left(8\lambda-9\mu\right)\lambda_h^2}\right\}\, .
\end{align}
Note that 
\begin{equation}\label{prodevalues}
2\Delta^2 E_1E_2 = - \mu 
\left(\lambda-\mu\right) \left[ \lambda^2 - \lambda_h^2 - \mu\lambda \right] \, . 
\end{equation}
The last factor  in this expression is necessarily positive for an interior fixed point, so the Poincar{\' e}  index is $-1$ for $\mu>0$ and $+1$ for $\mu<0$. Index considerations therefore allow limit cycles only for $\mu<0$. However, $\mu<0$ corresponds to $\Lambda<0$, and in this case  the constraint (\ref{ineq}) implies that $\varphi$ is a monotone function, so  limit cycles are excluded.

\subsection{Phase portraits}

We now present a selection of global phase portraits for representative values of the parameters. Grey dots indicate saddle points and black dots indicate nodes/foci. In each case the physical phase space is the interior, and boundary, of an ellipse. This allows for both signs of $\Lambda$, with $\Lambda<0$ in the upper and lower `lobes'. 

The  upper and lower segments of the  elliptical boundary could be identified, in which case the endpoints of the two curves `at infinity'  would be joined to create a continuous closed curve bounding  a connected  $\Lambda<0$ phase space with the topology of a disc. This is the phase space described for $\mu=0$ in \cite{Bergshoeff:2003vb}, where the disc was viewed as the `northern' hemisphere of a 2-sphere\footnote{The analysis of \cite{Bergshoeff:2003vb} was for $\Lambda>0$ cosmologies but this corresponds to $\Lambda<0$ domain walls. The `southern' hemisphere of the 2-sphere of this reference consisted of the time-reversed trajectories for the {\it same} sign of $\Lambda$.}.  However, a similar identification of the remaining two segments of the elliptical boundary (representing the constant axion trajectories for $\Lambda>0$) is not possible.

\newpage
\begin{figure}[ht!]
  \hfill
  \begin{minipage}[t]{.45\textwidth}
    \begin{center}  
      \epsfig{file=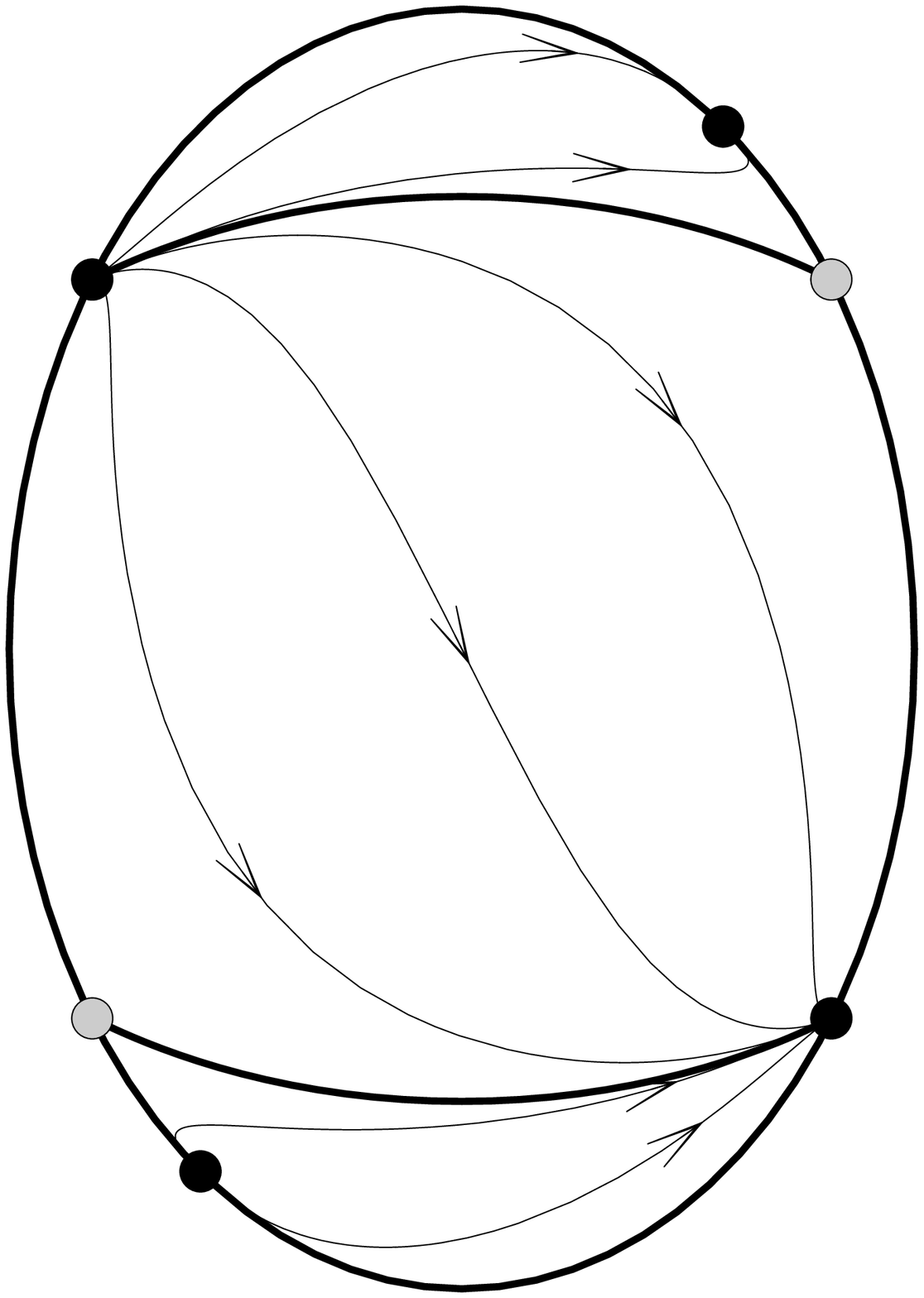, scale=0.38}
      \caption{{\footnotesize $\lambda < \lambda_h$; $\mu>0$. There are no fixed points in the interior of the allowed
      region. Saddles are grey dots, nodes black dots.}\label{fig:ellipse}}
    \end{center}
  \end{minipage}
  \hfill
  \begin{minipage}[t]{.45\textwidth}
    \begin{center}  
      \epsfig{file=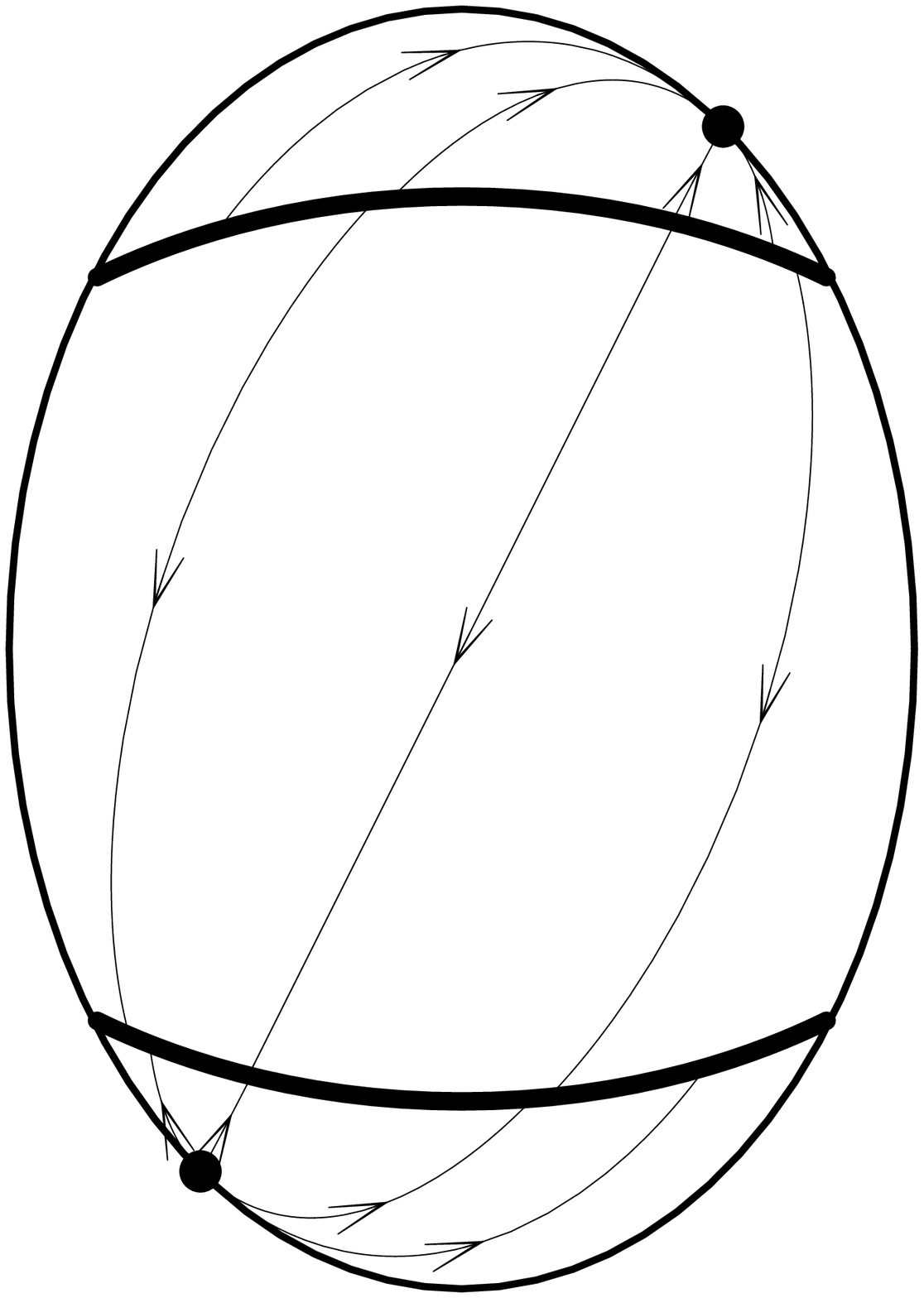, scale=0.38}
      \caption{{\footnotesize $\lambda<\lambda_h$; $\mu = 0$. The segments of the orthogonal ellipse
      at infinity turn into lines of fixed points on which there is no flow.}\label{fig:musiszerolambdalesslambdah}}
    \end{center}
  \end{minipage}
  \hfill
\end{figure}
\vfill
\begin{figure}[h!]
   \hfill
   \begin{minipage}[t]{.45\textwidth}
     \begin{center}  
       \epsfig{file=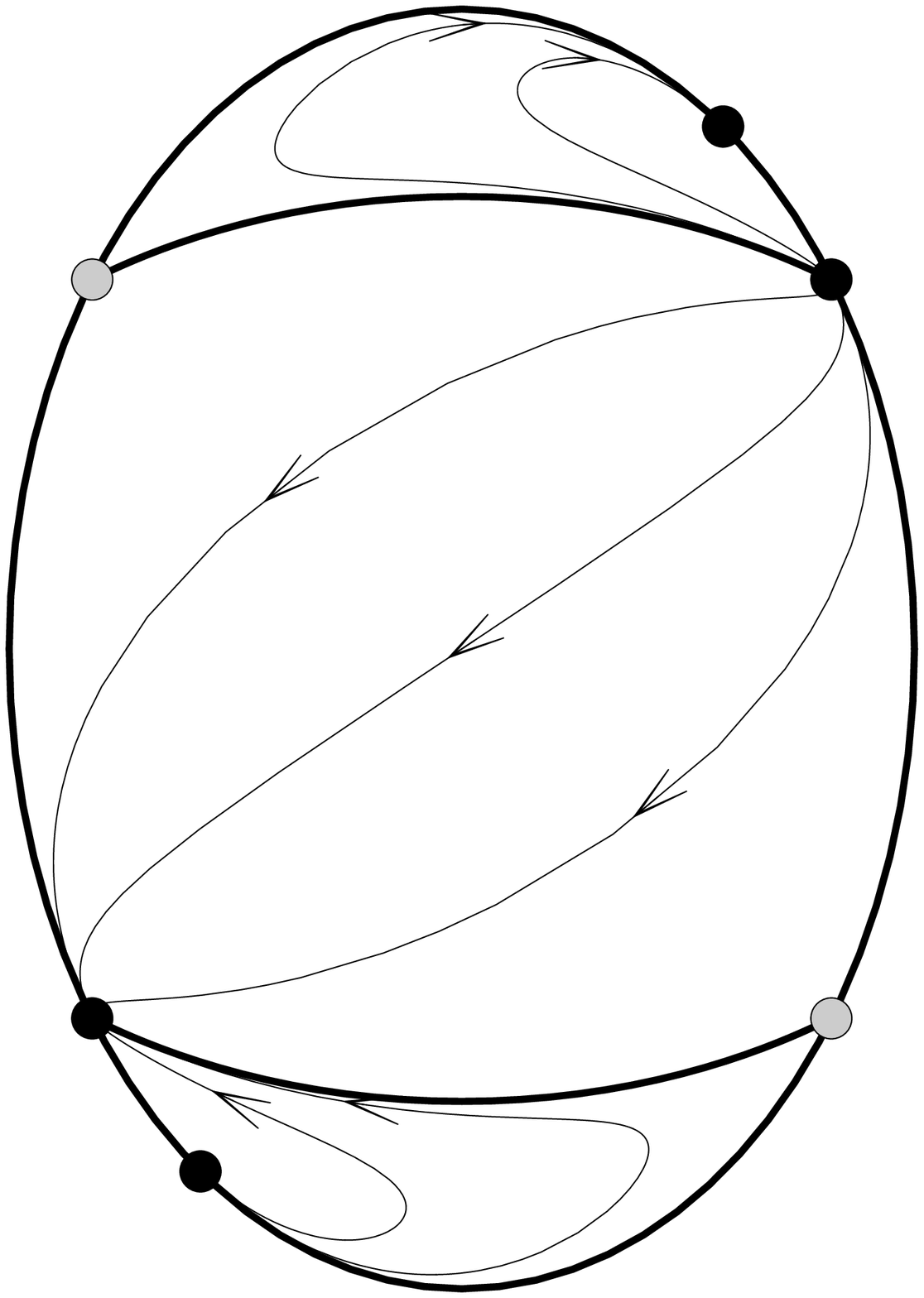, scale=0.38}
       \caption{{\footnotesize $\lambda<\lambda_h$; $\mu_c<\mu <0$.}\label{fig:ellipse2}}
  \end{center}
   \end{minipage}
  \hfill
   \begin{minipage}[t]{.45\textwidth}
     \begin{center}      
 	\epsfig{file=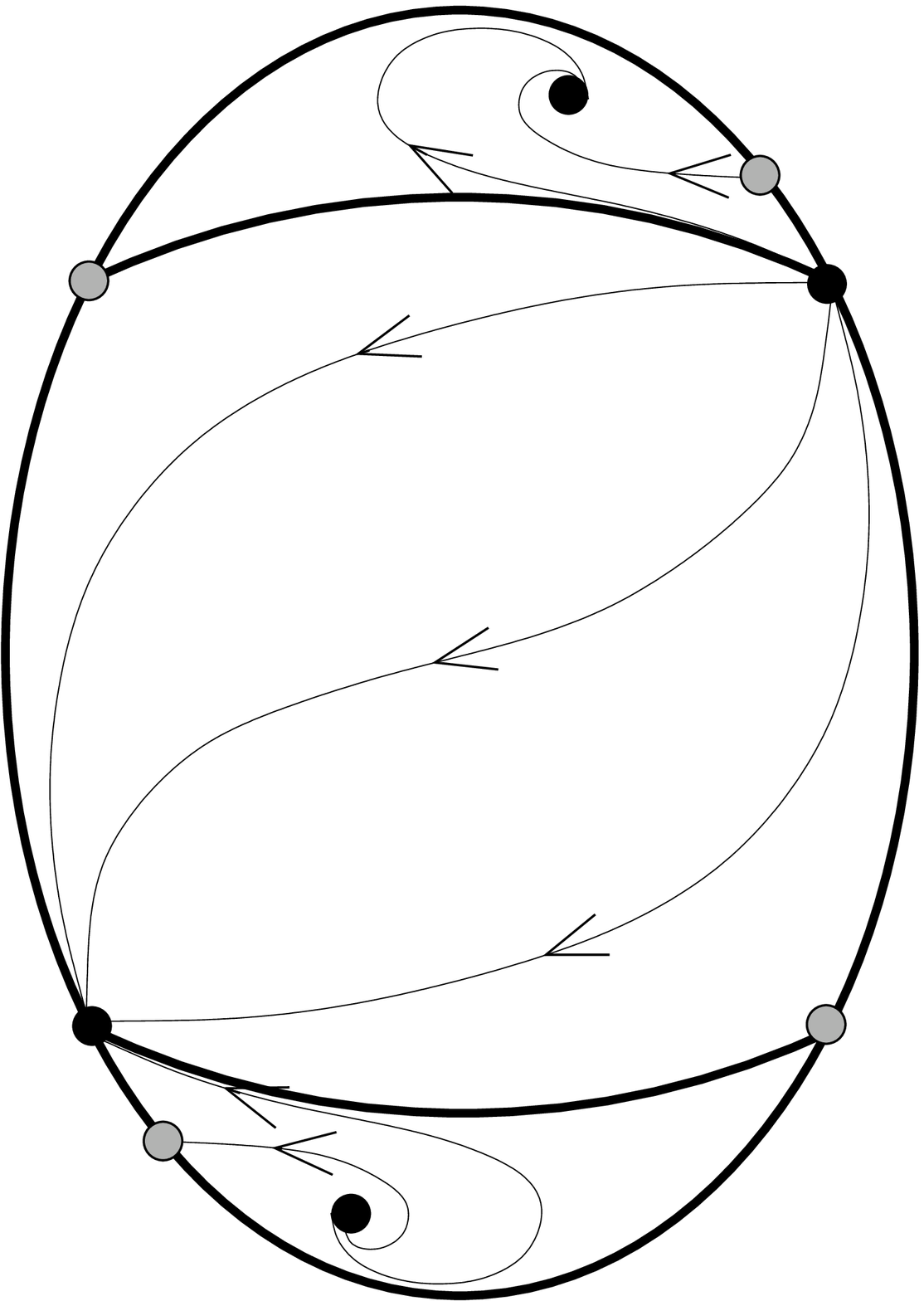, scale=0.36}
	\caption{{\footnotesize $\lambda<\lambda_h$; $\mu <\mu_c<0$. There is now a fixed point inside the physical region.}\label{fig:figure4}}
      \end{center}
   \end{minipage}
   \hfill
 \end{figure}
 \vfill
\newpage
\begin{figure}[h!]
  \hfill
  \begin{minipage}[t]{.45\textwidth}
    \begin{center}  
      \epsfig{file=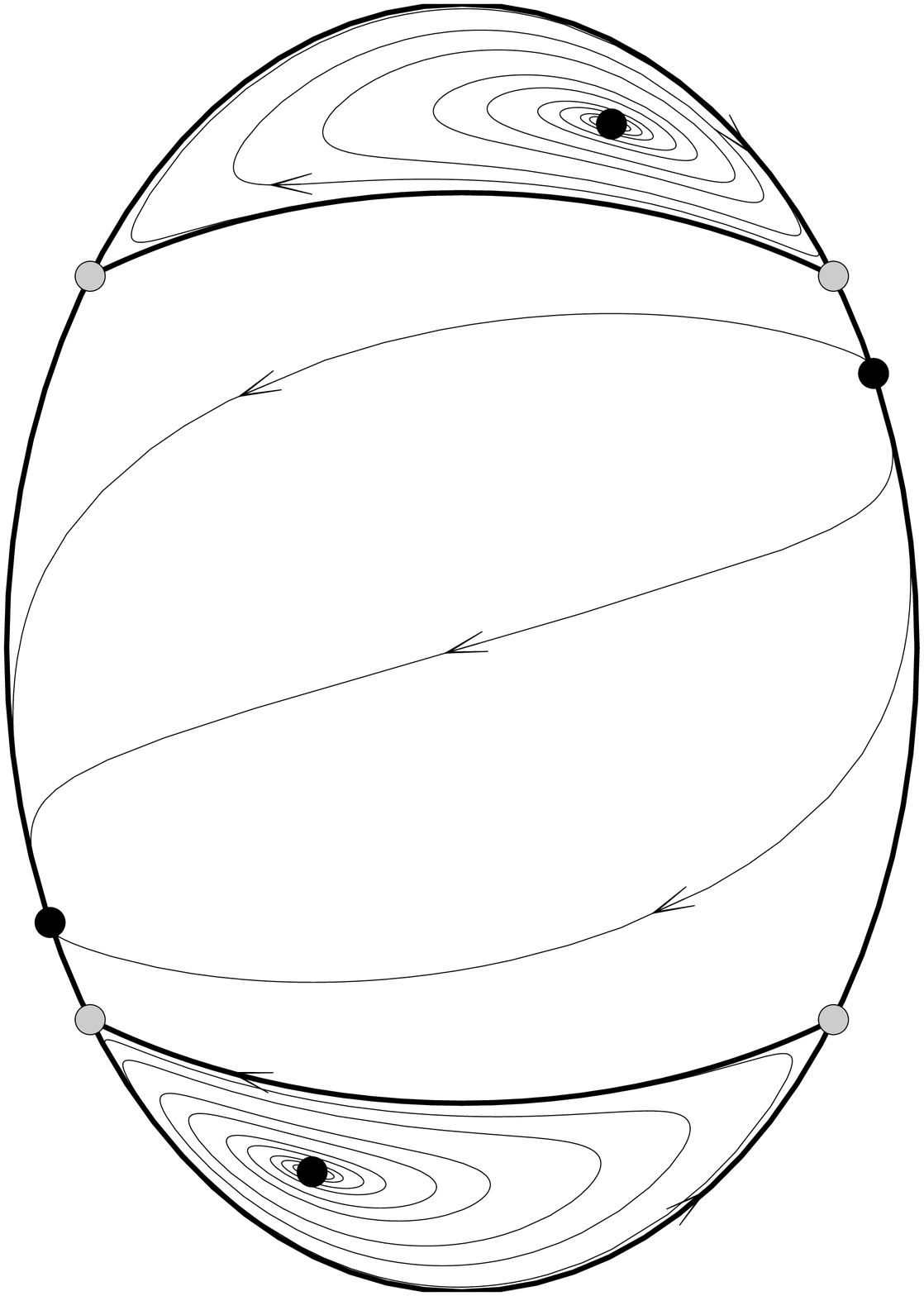, scale=0.3}
      \caption{{\footnotesize $\lambda > \lambda_h$; $\mu<0$. Note the quasi-cyclical behaviour in the $\Lambda<0$ regions.}\label{fig:cyclical}}
      \vfill
    \end{center}
  \end{minipage}
  \hfill
  \begin{minipage}[t]{.45\textwidth}
    \begin{center}  
      \epsfig{file=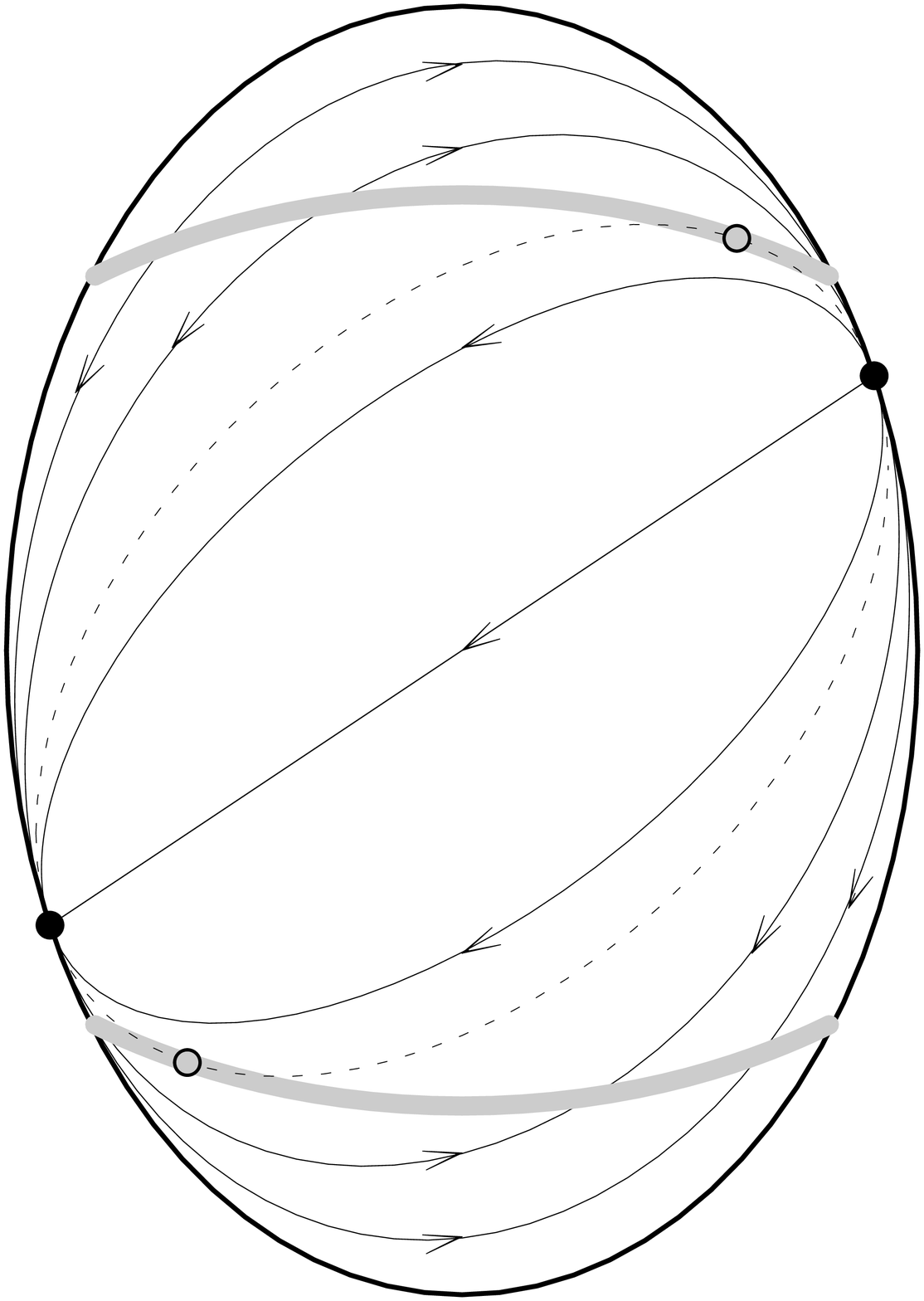, scale=0.27}
      \caption{{\footnotesize $\lambda>\lambda_h$; $\mu = 0$. The segments of the orthogonal ellipse
      at infinity turn into lines of fixed points on which there is no flow.}\label{fig:muiszerolambdaglambdah}}
      \vfill
    \end{center}
  \end{minipage}
  \hfill
\end{figure}
\vfill
\begin{figure}[h!]
  \hfill
  \begin{minipage}[t]{.45\textwidth}
    \begin{center}
      \epsfig{file=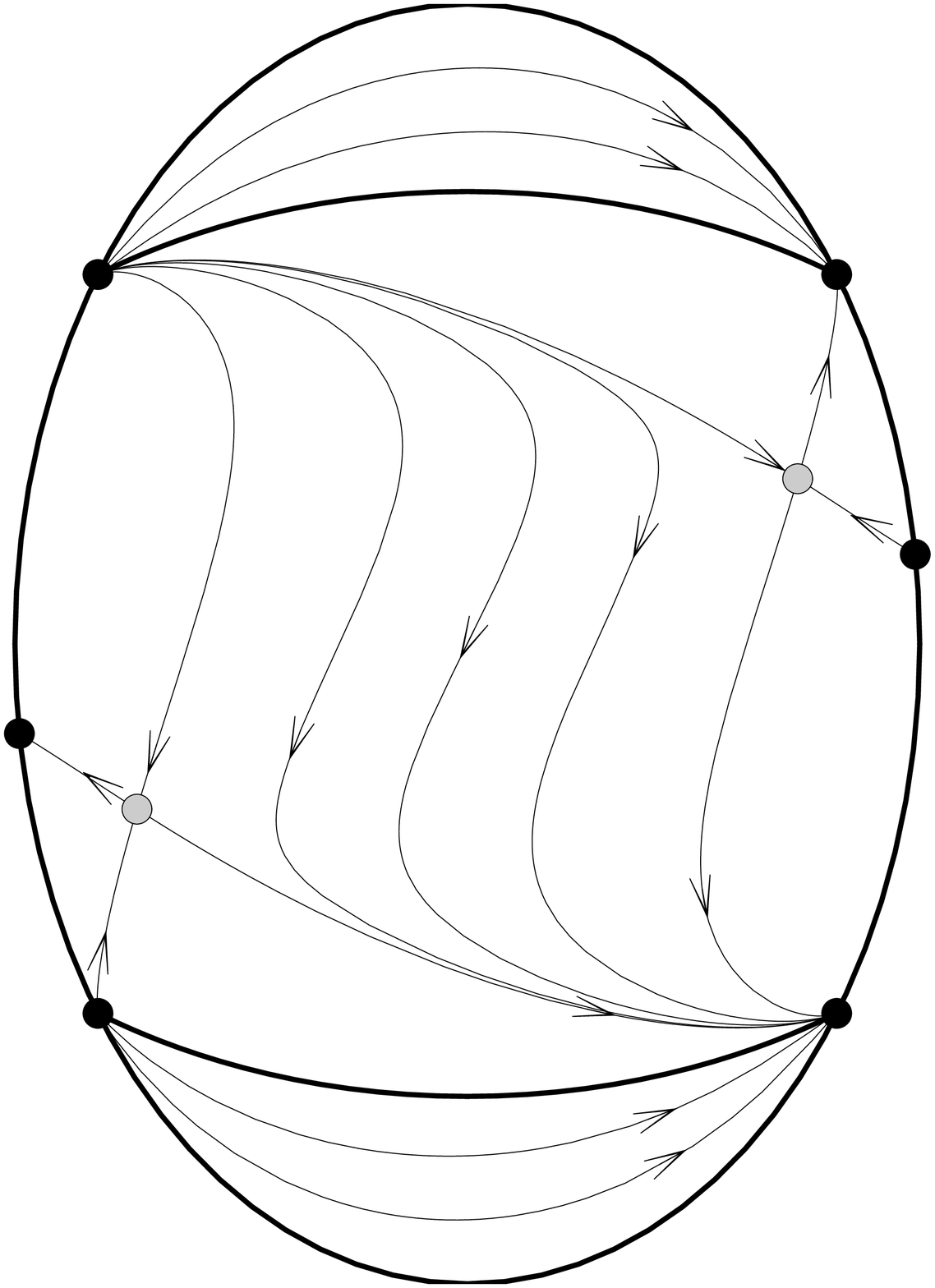, scale=0.3}
      \caption{{\footnotesize $\lambda>\lambda_h$; $\mu_c > \mu > 0$.}\label{fig:ellipselambda3}}
    \end{center}
  \end{minipage}
  \hfill
  \begin{minipage}[t]{.45\textwidth}
    \begin{center}  
      \epsfig{file=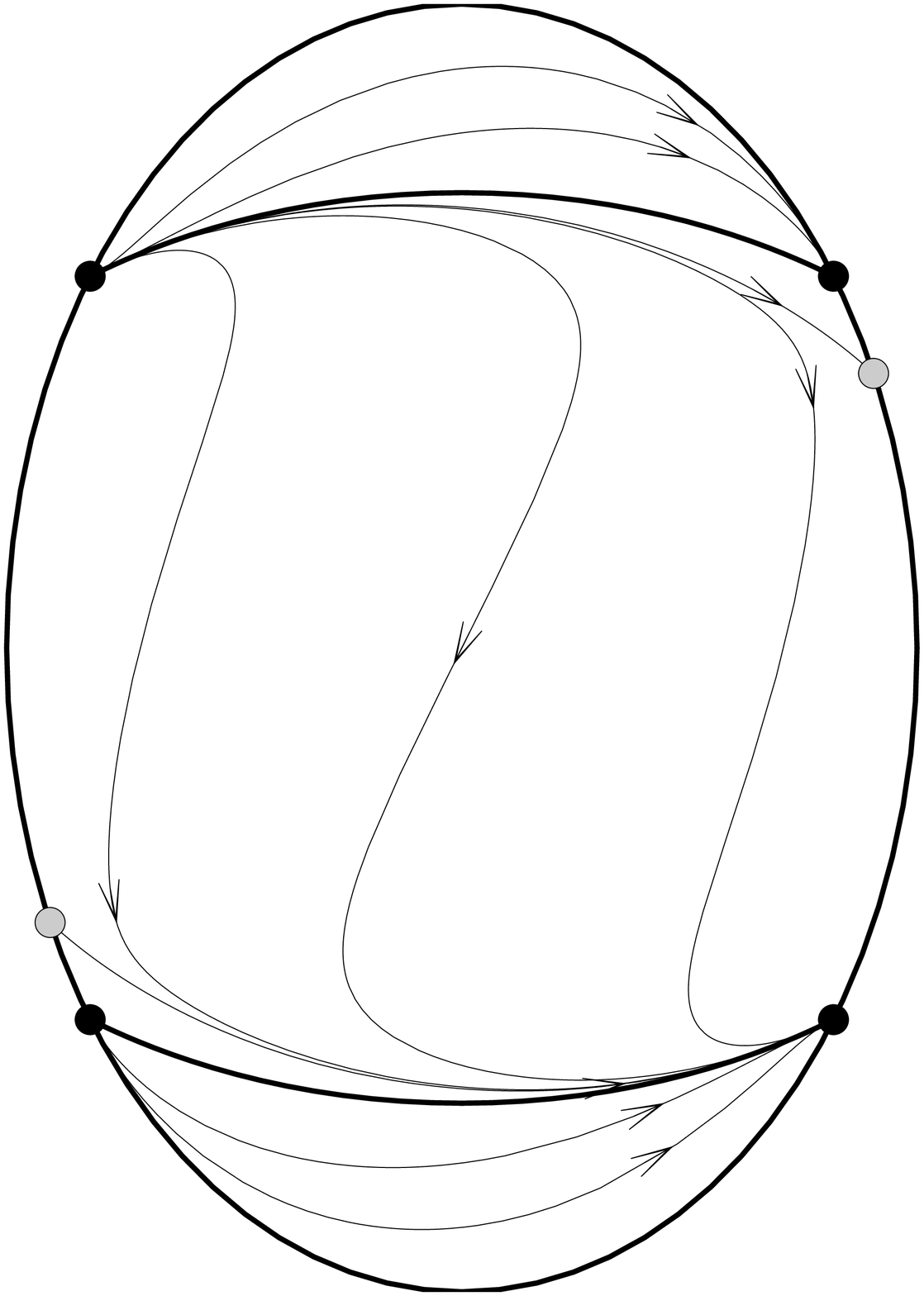, scale=0.3}
      \caption{{\footnotesize $\lambda>\lambda_h$; $\mu >\mu_c>0$.} }
    \end{center}
  \end{minipage}
  \hfill
\end{figure}
\vfill
\newpage

\subsection{Flat target space}

The $\mu=0$ case is special because the equations of motion then define a 3-dimensional autonomous
dynamical system in terms of the variables
\be
v =\dot \varphi \, , \qquad u= \dot\sigma \, , \qquad w= \dot\chi\, . 
\ee
These equations are
\begin{align}
\dot u &= -\alpha uv + \frac{1}{2}\lambda u^2 -\lambda\Lambda\, , \\
\dot w &= -\alpha wv + \frac{1}{2}\lambda u w\, \\
\dot v &= -\beta v^2 -\frac{1}{2\alpha} u^2 + \frac{1}{2}\lambda uv - 
\frac{1}{2\alpha}w^2 -2\beta\Lambda\, , 
\end{align}
and the  constraint is
\be
v^2-u^2-w^2 +2\Lambda = \frac{k}{\beta^2}e^{\lambda\sigma -2\beta\varphi}\, . 
\ee
For $w=0$ we recover the equations of (\ref{DS1}) and the constraint (\ref{constraint}) of the constant axion case. Moreover, fixed points of this sub-system are also fixed points of the 3-dimensional dynamical system.  

For $k=0$, we may use the constraint to eliminate the variable $w$ from the equations for $(\dot u,\dot v)$, which then become those of (\ref{DS2}), while the equation for $\dot w$ becomes redundant because it follows from differentiation of the $k=0$ constraint.  However, fixed points of this $k=0$ `subsystem' are not generally fixed points of the 3-dimensional dynamical system because there are no $w\ne0$ fixed points of the latter unless $\lambda=0$ and $\Lambda<0$ (in this special case there is circle of fixed points with $v=0$ and $u^2+w^2= 4\alpha\beta|\Lambda|$, but these are $k\ne0$ fixed points). What this means is that the fixed points of the 2-dimensional $k=0$ `subsystem' correspond to non-trivial $k=0$ trajectories of the 3-dimensional system. 

It is instructive to consider the special case of $\lambda=0$. If we write
\be
u+iw = R\, e^{i\psi}
\ee
then the equations of the 3-dimensional system reduce to the one equation $\dot\psi=0$ and the following two equations:
\be
\dot R = -\alpha  R v  \, , \qquad 
\dot v = -\alpha v^2 + {1\over2 \alpha} R^2 -2\beta \Lambda\, . 
\ee
These are just the constant axion equations  (\ref{DS1}) at $\lambda=0$ with $u\to R$, except that 
$R$ cannot be negative. The demonstration of section \ref{sec:constantchi} that the constant axion  phase space is globally a sphere, when allowance is made for both signs of $\Lambda$, now leads to the conclusion that the global phase space is a 3-sphere. This conclusion also holds when $\lambda\ne0$, in which case $\dot\psi\propto \sin\psi$, and the union of the $\psi=0$ and $\psi=\pi$ subspaces may be identified with the  spherical phase space of the constant axion system.

\subsection{Bifurcations}

\begin{figure}[t]
\begin{center}
 \epsfig{file=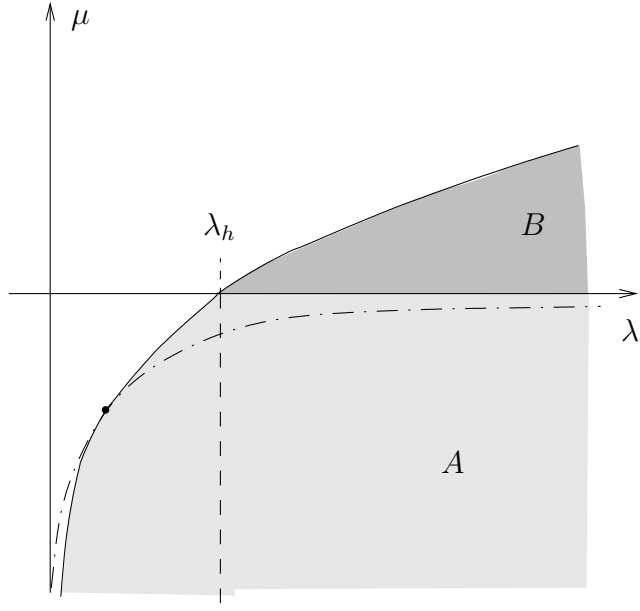, scale=0.7}
 \begin{picture}(0,0)(0.01,0.01)
	\put(-220,220){$\mu$}
	\put(-12,100){$\lambda$}
	\put(-170,140){$\lambda_h$}
	\put(-80,50){$A$}
	\put(-50,140){$B$}
	\end{picture}
\caption{{\footnotesize The $(\mu,\lambda)$ plane. For parameter values inside the semi-infinite grey region $A\cup B$ there exist internal fixed points. The curve bounding this region from above is given by the equation $\lambda\mu = \lambda^2 - \lambda_h^2$, i.e. $\mu=\mu_c$. For $\mu<0$ (A) these are located in the two $\Lambda<0$ regions and for $\mu>0$ (B) they are in the $\Lambda>0$ region. The dash-dotted lines indicate the curve $\mu\lambda=-2$  in relation to the curve $\mu=\mu_c$ for $d=4$.}\label{fig:mudiag}}
\end{center}
\end{figure}

As shown by (\ref{prodevalues}), the pair of `internal' fixed points are non-hyperbolic when $\mu\lambda = \lambda^2-\lambda_h^2$, which is equivalent to $\mu=\mu_c$.  Recall that $\mu\le\mu_c$ was the condition for the `internal'  fixed points  to lie within the allowed region, so when $\mu=\mu_c$ they have moved to the boundary of the allowed region, where they coincide with the boundary fixed points of (\ref{secondpair}). Here we study the nature of the bifurcation as the curve $\mu= \mu_c(\lambda)$ is  crossed in the $(\lambda,\mu)$ plane (Fig. \ref{fig:mudiag}).

Without loss of generality we may concentrate on the pair of fixed
point in the $Y>0$ region, and we define new variables $(\breve X,\breve Y)$ by
\be
X = \frac{\lambda}{\sqrt{\lambda_h^2+2\lambda^2}}
-\frac{\lambda_h}{2\lambda} \breve X - \frac{\lambda^2 +
  2\lambda_h^2}{3\lambda\lambda_h} \breve Y \, , \qquad
Y=  \frac{\lambda_h}{\sqrt{\lambda_h^2 + 2\lambda^2}} + \breve X + \breve Y\, . 
\ee
The fixed point is now at the origin in the $(\breve X,\breve Y)$ plane. We are
interested in the behaviour for $\mu \approx \mu_c(\lambda)$, so we
also define
\be
s = \frac{\lambda\left(\mu-\mu_c\right)}{\sqrt{\lambda_h^2 +
    2\lambda^2}}\, .
\ee
The equations of the dynamical system near the origin in
$(\breve X,\breve Y,\nu)$ space take the form
\be
\frac{d\breve X}{d\zeta} =
\frac{\lambda^2-\lambda_h^2}{2\sqrt{\lambda_h^2+2\lambda^2}}\breve X +
Q_1\left(\breve X,\breve Y;s\right) \, , \qquad
\frac{d\breve Y}{d\zeta} = s\,  \breve Y + Q_2\left(\breve X,\breve Y;s \right)\, , 
\ee
together with the trivial equation $s'=0$. The functions $Q_1$ and
$Q_2$ are polynomials in $(\breve X,\breve Y)$ that are at least quadratic in
these  variables, with coefficients that may depend on $s$. For our 
purposes, the relevent terms are
\be
Q_1= \frac{13\lambda_h^2-\lambda^2}{9\lambda_h} \breve Y^2 +
\cdots\, , \qquad
Q_2 = -\frac{2\lambda_h\left(\lambda_h^2 +
  2\lambda^2\right)}{3\lambda^2} \breve Y^2 + \cdots
\ee
The equation for the centre manifold of this system (i.e. the
`extended' centre manifold) is, to quadratic order,  
\be
\breve X  =
-\frac{2\sqrt{\lambda_h^2+2\lambda^2}\left(13\lambda_h^2-
\lambda^2\right)}{9\lambda_h\left(\lambda^2-\lambda_h^2\right)}
\breve Y^2 + \cdots
\ee
Evidently, we must assume that $\lambda\ne\lambda_h$, since in this
case there is a coincidence of three fixed points and the bifurcation 
is more complicated. The dynamics for the evolution on the extended centre
manifold is governed by the equation
\be
\frac{d\breve Y}{d\zeta}= s\, \breve Y -
\frac{2\lambda_h\left(\lambda_h^2+2\lambda^2\right)}{3\lambda^2}
\breve Y^2 + \cdots
\ee
From this formula it is evident that the bifurcation is transcritical. 

There is a much more complicated bifurcation at $\mu=0$. Again we
restrict to $\lambda \neq\lambda_h$. In this case, equations (\ref{eq:XYsystem}) imply that the two segments of the ellipse $X^2 + 2Y^2 =1$ inside the physical region become lines of fixed points. The shape of the trajectories follows from the differential equation
\begin{equation}
\frac{d Y}{d X} = \frac{2\lambda XY- \lambda_h\left(1-Y^2\right)}{\lambda_h XY -\lambda\left(1-2X^2\right)}\, , 
\end{equation}
which has the solution
\begin{equation}\label{eq:ellipsemuiszero}
c (\lambda Y - \lambda_h X)^2 = 1-Y^2 - 2 X^2\,,
\end{equation}
for constant $c$. Trajectories inside the physical region have  $c\geq 0$,  and $c=0$ 
corresponds to the boundary ellipse. All trajectories are segments of some ellipse 
that passes through the fixed points on the intersection of the boundary with the 
line $\lambda_h X = \lambda Y$. These fixed points are in the $\Lambda<0$ regions
if $\lambda<\lambda_h$ (Fig. \ref{fig:musiszerolambdalesslambdah}) and
in the $\Lambda>0$ region if $\lambda>\lambda_h$
(Fig. \ref{fig:muiszerolambdaglambdah}). If $\lambda>\lambda_h$, then
$\mu_c(\lambda)>0$, so that $\mu<\mu_c$ at the bifurcation. In this case
each $\Lambda<0$ region is bounded by a heteroclinic cycle, and this leads to the
 quasi-cyclical $\Lambda<0$ trajectories of Fig.  \ref{fig:cyclical}.  

As $\mu$ approaches zero, the interior fixed point  approaches the line at infinity and 
merges with it at $\mu=0$; there  this line turns into a line of fixed points. The physical trajectories 
are segments of ellipses (\ref{eq:ellipsemuiszero}) emanating from the 
boundary fixed points on either side of the $\dot\chi=0$ boundary 
of the $\Lambda>0$ region (Fig. \ref{fig:muiszerolambdaglambdah}). 
There is a unique ellipse, with $1/c = \lambda^2 - \lambda_h^2$, which 
just grazes the two boundaries at infinity (indicated by a dashed 
line in Fig. \ref{fig:muiszerolambdaglambdah}); it touches them
precisely at the limiting position of the interior fixed point as 
$\mu\rightarrow 0$. Furthermore, the location of this fixed point at 
$\mu=0$ divides the lines at infinity into two regions. Concentrating 
on the $Y>0$ case, we have a line of repellers to the left of the 
fixed point and a line of attractors to the right, making this line 
a kind of extended saddle.  As $\mu$ becomes negative, the interior 
fixed point wanders into the $\Lambda>0$ region where it becomes a 
proper saddle. The drastic change in the topology of the flow is 
illustrated in Figs. \ref{fig:cyclical},
\ref{fig:muiszerolambdaglambdah} and \ref{fig:ellipselambda3}.
If $\lambda<\lambda_h$, there is no interior fixed point in the
physical region as $\mu$ goes through zero. The corresponding 
bifurcation is illustrated in Figs. \ref{fig:ellipse} -
\ref{fig:ellipse2}. At $\mu=0$ the trajectories are again given by
segments of ellipses (\ref{eq:ellipsemuiszero}), now emanating from 
the boundary fixed points in the $\Lambda<0$ regions. The two lines 
`at infinity' are again lines of fixed points. There is now a
universal repeller in the $Y>0$ region and a universal attractor 
in the $Y<0$ region. 

\subsection{An exact axion-dilaton domain-wall solution}
\label{sec:exact}

The interior fixed point solution has
\be
(u,v) = \pm A \left(\lambda_h,\lambda-\mu\right)\, ,  \qquad
e^{\frac{1}{2}\mu\sigma} \dot\chi = \left(\frac{\mu\lambda_h}{2}\right) \kappa\, A
\ee
where 
\be\label{Akappa}
A= \sqrt{\frac{2\Lambda}{\mu\left(\lambda-\mu\right)}}\, , \qquad
|\kappa| = \frac{2}{\lambda_h} \sqrt{\frac{\lambda\left(\mu_c-\mu\right)}{\mu^2}}\, . 
\ee
Recall that the existence of an interior fixed point requires that both $\lambda-\mu$ and $\Lambda/\mu$ be positive, and that $\mu<\mu_c$, so that the constants $A$ and $\kappa$ are real. 

Integrating  the equations $\dot \sigma = u$ and $\dot v = \varphi$,  we obtain the following  new exact domain-wall solution
\be\label{fixedinterior}
\sigma = \pm \lambda_h A (z-z_0)\, , \qquad 
\varphi =  \pm \left(\lambda-\mu\right)A(z-z_0) + \varphi_0\, , 
\ee
where $z_0$ and $\varphi_0$ are integration constants, and $\chi$ is such that
\be\label{sigchi}
e^{{1\over2}\mu\sigma}\left(\chi-\chi_0\right) = \kappa\, ,
\ee
where $\chi_0$ is a further integration constant.

The spacetime metric for a convenient choice of $\varphi_0$ is
\be
ds^2= d\rho^2 + \rho^{4\beta/\lambda} ds^2\left(\bE^{(1,d-2)}\right)\, ,
\ee
where
\be
\rho = \pm \frac{2}{\lambda\lambda_h A} \exp\left(\pm \frac{\lambda\lambda_h}{2} A (z-z_0)\right)\, . 
\ee
This solution will play an important role in the discussion of supersymmetry to follow.  

Although the limiting case in which $\mu=\mu_c$ yields a solution with constant axion, this case is of interest because the $\mu=\mu_c$ models include the $d=4$ Freedman-Schwarz (FS) supergravity theory, which has a supersymmetric domain wall solution \cite{Cowdall:1997fn}. In fact, the above solution generalizes the FS domain-wall to any $d$ when $\mu=\mu_c$. More significantly, it shows that the further generalization to $\mu<\mu_c$ requires a non-constant axion field.

\section{Supersymmetry}
\setcounter{equation}{0}

For domain wall solutions of supergravity theories it is of interest to ask of any given solution whether it is supersymmetric, since supersymmetry implies stability.  A necessary condition for a domain wall solution to be supersymmetric is that it admit a Killing spinor, and this condition is also sufficient for models with only a single scalar field, e.g. a dilaton field, because the vanishing of the dilatino supersymmetry variation is then an integrability condition for the existence of a Killing spinor.  Another simplifying feature of single-scalar models is that both the potential $V$ and the Killing spinor equation are determined by a `superpotential'  alone; in the multi-scalar case there is an additional dependence on the target space metric. An example of a single scalar supergravity model  is minimal $d=3$ supergravity coupled to a scalar multiplet. In this case the superpotential is a real function $W(\sigma)$ of the single scalar field $\sigma$ and the potential is given in terms of $W$ by the formula
\be\label{VW}
V= 2\left[\left(W'\right)^2-\alpha^2W^2\right]\, , 
\ee
where the prime indicates a derivative with respect to $\sigma$. This formula can be immediately generalized to $d$ dimensions, as is implicit in the way it has been written. The analogous $d$-dimensional generalization of the Killing spinor equation is 
\be\label{Kill2}
\left[D_\mu - \frac{1}{2(d-2)} W \,\Gamma_\mu \right]\epsilon =0\, ,
\ee
where $D_\mu$ is the usual covariant derivative acting on spinors. 
In the $d=3$ supergravity context this equation arises from the requirement of  vanishing supersymmetry variation of the gravitino field, in which case $\epsilon$ is an anticommuting spinor parameter,  but linearity of the equation allows us to re-interpret it as a commuting spinor  field, now on a $d$-dimensional spacetime. 

The point of this generalization to arbitrary $d$ of the notion of a supersymmetric domain wall solution of $d=3$ supergravity is that domain walls admitting Killing spinors with respect to some superpotential $W$ will be classically stable even when there is no underlying supergravity theory, although stability against non-perturbative tunnelling to some lower-energy configuration  is not precluded. This state of affairs is described by saying that $W$ defines a `fake' supergravity theory.  Its `fake' supersymmetric domain wall solutions satisfy 
\be\label{susywithf}
f^{-1}\dot\varphi = \mp 2\alpha e^{\alpha\varphi} W\, , \qquad 
f^{-1}\dot\sigma= \pm 2e^{\alpha\varphi}W'\, ,  
\ee
because any solution of these equations is a flat domain wall solution of the second-order equations for $(\varphi,\sigma)$, and of the diffeomorphism constraint, such that the domain wall metric admits a Killing spinor.  A fake supersymmetric domain-wall may be genuinely supersymmetric if the superpotential arises in the context of some `genuine' supergravity theory with the required bosonic truncation, such that (i) the potential takes the form (\ref{VW}) and (ii) the condition of vanishing supersymmetry variation of the gravitino field reduces to (\ref{Kill2}). The conditions under which this happens for $d=5$ supergravity were investigated in \cite{Celi:2004st}, and we will determine here the conditions under which this happens for $d=4$ supergravity. When there is an underlying supergravity theory we shall say that the domain wall solution is `genuinely' supersymmetric, and we will usually omit the adjective `fake' that should qualify the generic case because this should be clear from the context. 

Only special solutions can be supersymmetric with respect to a {\it given} superpotential $W$, but almost any flat domain wall determines a superpotential $W$ with respect to which it is supersymmetric \cite{Freedman:2003ax,Sonner:2005sj,Skenderis:2006jq}. The proof is by construction of $W$. Given $\varphi(z)$ and a choice of the function $f$, the first of equations (\ref{susywithf}) determines $W$ as a function of $z$. Then, given $\sigma(z)$, the inverse  function $z(\sigma)$ yields $W$ as a function of $\sigma$, and one may verify that $W(\sigma)$ so defined  yields the potential according to the formula (\ref{VW}).  A complex superpotential is needed for the same statement to be true of curved walls, but we shall restrict our attention to flat domain walls, for  which a real superpotential suffices.   The qualification `almost' arises because the function $z(\sigma)$ is defined only when $\sigma(z)$ is strictly monotonic, i.e. only when $\dot\sigma\ne0$. This  monotonicity condition fails for  adS vacua, viewed as domain walls, because these have $\sigma\equiv0$; this is expected because adS vacua may be unstable. It also fails for domain walls that are asymptotic to an unstable adS vacuum because in this case the asymptotic solution is an accumulation point for zeros of $\dot\sigma$ \cite{Skenderis:2006rr}. 

We will begin our investigation of domain wall supersymmetry by examining in more detail what happens at an isolated zero of $\dot\sigma$. We will confirm the `piecewise supersymmetry' conclusion of \cite{Skenderis:2006jq}  but the details are instructive and suggest a general picture that we elaborate in the context of  asymptotically adS domain walls.  Then  we  turn to a consideration of supersymmetry in multi-scalar models, using the axion-dilaton models as a `laboratory'. The exact solution of subsection \ref{sec:exact} is useful in this respect and what we learn from it motivates a new multi-scalar extension of the ideas of fake supergravity.  

\subsection{Single-scalar models revisited}

The formula (\ref{VW}) implies that
\be
V'= 4W'\left(W'{}'-\alpha^2 W\right)\, . 
\ee
It is usually concluded that stationary points of $W$ are stationary points of $V$, with $V<0$,
but this assumes that $W'{}'$ is non-singular when $W'=0$. It may happen that a zero of $W'$ is cancelled by a pole of $W'{}'$ such that $V'$ is finite and non-zero. We shall argue here that this possibility is realized  for any superpotential constructed from a solution for which $\dot\sigma$ has an isolated zero. Rather than present a general proof we illustrate the point with two examples, one explicit and the other implicit. The explicit example  arises from an exact domain-wall solution of the $\lambda=\lambda_h$ Einstein-dilaton model, for which $\dot\sigma$ has a single isolated zero \cite{Sonner:2005sj}. The implicit example arises in the context of asymptotically adS domain walls. 

Before proceeding we pause to discuss some consequences of the choice of the function $f$ in (\ref{susywithf}). For the choice made in (\ref{gc}), which coincides with the choice made in \cite{Sonner:2005sj}, the first-order equations (\ref{susywithf}) become
\be\label{susyfixed}
\dot\varphi = \mp \lambda_h e^{\frac{1}{2}\lambda_h\sigma} W\, , \qquad
\dot\sigma = \pm 2 e^{\frac{1}{2}\lambda_h\sigma} W' \, .
\ee
The choice $f= e^{-\alpha\varphi}$, which was made in \cite{Skenderis:2006jq}, leads to simpler first-order equations in terms of an affine distance variable, which we called $\tilde z$ in subsection 
\ref{sec:prelim}. In terms of this affine distance variable, defined as a function of $z$ by (\ref{affinez}) 
the equations (\ref{susyfixed}) become
\be\label{newsusy}
\grave\varphi = \lambda_h W \, , \qquad \grave\sigma =- 2W' \, . 
\ee
where the grave accent indicates differentiation with respect to $\tilde z$.

\subsubsection{Branched superpotentials}

 Let us first return to  the domain wall solution of the $\lambda=\lambda_h$ constant axion model with $\Lambda=-2/\lambda_h^2$ studied in subsection \ref{sec:traj}. Recall that $\dot\sigma(z)$ has a zero at $z=1$ for this solution.  Our aim is to elucidate the implication of this fact for supersymmetry. Using the solution to construct a superpotential in the way just reviewed, one finds that
 \be\label{simplesuperpot}
 W(\sigma)= \mp \left(\frac{1+z^2}{\lambda_h^2 z}\right) 
 e^{-\frac{1}{2}\lambda_h\sigma}\, , 
 \ee
where $z(\sigma)$ is determined implicitly by
\be
e^{\lambda_h\sigma} = z^{-1} e^{\frac{1}{2} z^2}\, . 
\ee
Observe that $z'= \lambda_h z/(z^2-1)$, and hence that
\be
W' = \mp\left( \frac{1-z^2}{2\lambda_h z}\right)e^{-\frac{1}{2}\lambda_h\sigma}\, . 
\ee
We see that $W'$ vanishes at $z=1$, i.e. at $\sigma=\sigma_{min}\equiv \sigma(1)$. One the one hand, this is required by the second of equations (\ref{susyfixed}), since $\dot\sigma=0$ at $z=1$. On the other hand, the absence of any stationary points of $V$ would lead us to expect the same of $W$. This apparent contradiction is resolved by  the observation that 
\be
W'{}'-\alpha^2 W = \mp\left(\frac{z}{1-z^2} \right)
e^{-\frac{1}{2}\lambda_h\sigma}\, . 
\ee
This diverges precisely when $W'=0$, such that
\be
4W'\left[W'{}'-\alpha^2 W\right] \equiv \frac{2}{\lambda_h} e^{-\lambda_h\sigma} \equiv V'\, ,  
\ee
as expected. This illustrates the important point that {\it stationary points of $W$ are not necessarily stationary points of $V$}. 

\begin{figure}[t]
\begin{center}
 \epsfig{file=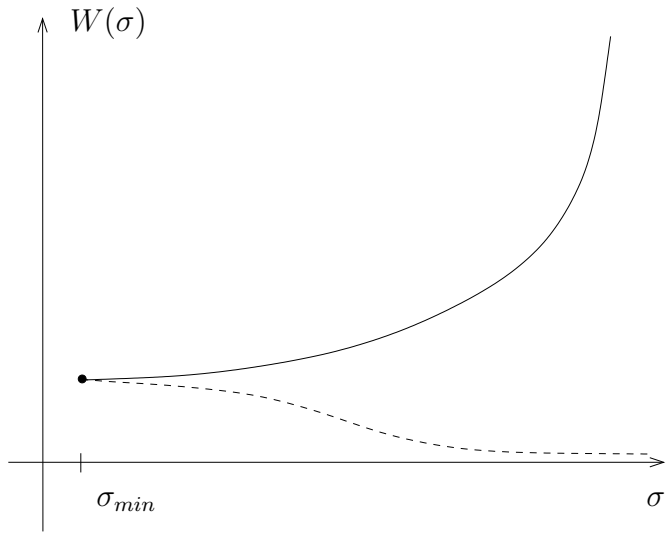, scale=0.46}
 \begin{picture}(0,0)(0.01,0.01)
	\put(-230,190){$W(\sigma)$}
	\put(-12,10){$\sigma$}
	\put(-220,10){$\sigma_{min}$}
	%\put(-80,50){$A$}
	%\put(-50,140){$B$}
	\end{picture}
\caption{{\footnotesize The branched superpotential. One branch is shown as a solid line, the other branch is shown as a dashed line. Both branches are defined for all $\sigma\geq\sigma_m$.}}
\end{center}
\end{figure}

The superpotential of (\ref{simplesuperpot}) is defined only for $\sigma>\sigma_{min}$.  This is not a problem {\it per se} because the solution is such that $\sigma\ge\sigma_{min}$, but the minimum of $\sigma(z)$ is a branch point of the function $z(\sigma)$. The same is therefore true of $W(\sigma)$. In fact, one has
\be\label{Wexp}
W(\sigma)\sim W_0 \left|\sigma-\sigma_{min}\right|^{\frac{3}{2}} \qquad {\rm as} \ \ 
\sigma\to\sigma_{min}
\ee
for constant $W_0$, implying that $W(\sigma)$ is  double-valued with a branch point at $\sigma=\sigma_{min}$. As $z$ varies through $1$, the function $W(\sigma)$ switches from one branch to the other. In other words, the solution is `piecewise supersymmetric'.

\subsubsection{Asymptotically adS walls}

We now move on to more general potentials $V(\sigma)$. Of particular interest, and common occurrence in supergravity models, are potentials with stationary  points of $V$ at which $V<0$. These yield adS vacua. Given such a stationary point at $\sigma=0$, the potential has the expansion 
\be\label{expandpot}
V= - \frac{1}{2\beta^2\ell^2} + \frac{1}{2} m^2\sigma^2 + \dots
\ee
where $\ell$ is the adS radius and $m$ the mass of the $\sigma$-particle in this vacuum. 
The vacuum will be perturbatively stable as long as $m^2$ satisfies the $d$-dimensional  version of the
Breitenlohner-Freedman (BF) bound \cite{Breitenlohner:1982bm,Mezincescu:1984ev}
\be\label{BF}
m^2 \ge -\frac{\left(d-1\right)^2}{4\ell^2} \equiv m^2_{BF}\, . 
\ee
For a domain wall (flat or curved) that is asymptotic to the adS vacuum as $\tilde z\to\infty$, one has \cite{Skenderis:2006rr}
\be\label{asymp}
\sigma \sim e^{-\nu \tilde z/\ell}\, , \qquad 
\grave \varphi \sim \frac{1}{\beta\ell} + \frac{\alpha\nu}{2\ell} e^{-2\nu\tilde z/\ell}\, ,
\ee
where $\nu$ is a real positive root of the quadratic function $\nu^2 -(d-1)\nu - m^2\ell^2$.  
The roots 
\be
\nu_\pm = \frac{1}{2}\left[d-1 \pm \sqrt{\left(d-1\right)^2 + 4m^2\ell^2}\right]\, , 
\ee
are real as long as the BF bound is satisfied, and if $m^2<0$ then both $\nu_+$ and $\nu_-$ are 
positive.  If $m^2\ge0$ then only $\nu_+$ is positive.  

If we use the asymptotic solution (\ref{asymp}) for $\nu=\nu_\pm$ to construct the superpotential with respect to which it is supersymmetric then we may assume that $W\ge0$ at $\sigma=0$ without loss of generality because the overall sign of $W$ serves only to distinguish between walls and `anti-walls'. 
One finds the superpotential
\be\label{Wplusminus}
W_\pm = \frac{1}{2\alpha\beta\ell} + \frac{\nu_\pm}{4\ell} \sigma^2 + \dots
\ee
The construction guarantees that this solves (\ref{VW}), as one may verify, although we actually find the $W_-$ superpotential this way only for $m^2>0$ because for $m^2\le0$ there is no domain wall associated with $W_-$ because $\nu_-$ is not positive\footnote{This generalizes an observation in 
\cite{Sonner:2005sj} for the $m^2=0$ case of constant $V$: in that case $W_-=W_0$ and $W_+= W_0\cosh \left(\alpha\sigma\right)$ for constant $W_0$ but whereas both $W_+$ and $W_-$ admit supersymmetric adS vacua only $W_+$ admits a supersymmetric domain wall.}.  As noted in 
\cite{Freedman:2003ax}, in a somewhat different context, there is a further one-parameter family of perturbative solutions of (\ref{VW}) for which $W'(0)$ is non-zero, and a domain wall that is supersymmetric with respect to such a superpotential is {\it not} asymptotic to the adS vacuum at $\sigma=0$. These `other' superpotentials, $W_M(\sigma)$, parametrized by the non-zero real number $M$,  have the expansion
\begin{align}
W_M(\sigma) =& \sqrt{M^2+ \frac{(d-2)^2}{\ell^2}} + \alpha M\sigma + \frac{1}{2}\alpha^2 \sqrt{M^2+ \frac{(d-2)^2}{\ell^2}}\, \sigma^2 \nonumber \\
&+ \frac{1}{6}\left(\alpha^2 M^2 + \frac{m^2}{4\alpha M}\right)\sigma^3 + \dots
\end{align}
For non-zero $m$, it  is evident from the $\sigma^3$ term why $M$ must be non-zero. For $m=0$ we require $M\ne0$ because the $M=0$ solution then coincides with the $W_-$ superpotential. 

To summarize, each perturbatively-stable adS vacuum is associated with two superpotentials $W_\pm$ that are stationary at the vacuum and a family of superpotentials $W_M$ that are not. The question that we wish to address now is how the superpotentials defined at a maximum of $V$ are related to those defined at an `adjacent' minimum of $V$. A number of single-scalar supergravity models have a potential with a single adS maximum and a single adS minimum, derived from a superpotential
with respect to which the maximum is supersymmetric and the minimum is not. The expansion of this superpotential about the minimum necessarily coincides with $W_M$ for some value of $M$, and it appears from inspection of a few cases that it should be identified with $W_-$ at the maximum. 
In any case, any supergravity model automatically gives us at least one superpotential that is defined as a single-valued function for all values of the scalar field and hence for all values in the interval between the maximum and the minimum. Let us assume that we are given one such superpotential, not necessarily arising from a supergravity model, and ask about the other superpotentials. There are evidently various possibilities. It could be that each superpotential that is perturbatively defined at the maximum of $V$ continues to one that is perturbatively defined at the minimum. Alternatively, it could be that  a pair of superpotentials defined perturbatively at, say, the maximum are actually two branches of a 
double-valued potential that is not defined at the minimum. It follows from recent work of Amsel et al. \cite{Amsel:2007im} that there exist potentials for which  this latter possibility is realized. We now present a version of their argument adapted to our needs. 

We assume (i) that $V(\sigma)$ has an adS maximum, at $\sigma=0$, and an adS minimum at $\sigma=\sigma_+$ but  no other stationary points, (ii) that the continuation of $W_+$, perturbatively defined at the origin, leads to a function that is single valued in the interval $[0,\sigma_+]$, (ii) that this function is stationary at $\sigma=\sigma_+$  What can we say under these circumstances about the continuation of the function $W_-$ defined at the origin? Both $\nu_+$ and $\nu_-$ are real and positive at the origin, by hypothesis, so not  only are both $W_+$ and $W_-$ positive near $\sigma=0$  but so also are their derivatives $W_+'$ and $W_-'$, as one sees from (\ref{Wplusminus}).  Writing 
(\ref{VW}) as
\be\label{VWroot}
\sqrt{2}\,  W' = \sqrt{V+ 2\alpha^2 W^2}\, ,  
\ee
we see that it remains true that $W_+>W_-$ as we increase $\sigma$ from zero, until a zero of either $W_+$ or $W_-$ is reached. Since the function $V+ 2\alpha^2 W_+^2$ is strictly greater than 
the function $V+2\alpha^2W_-^2$, and since the former is zero at $\sigma=\sigma_+$, there is necessarily a zero of $W_-'$ at some $\sigma_-<\sigma_+$. A priori,  this might be another  stationary point of $V$ (which would  have to be some solution of $W_+'{}'=\alpha^2W_+$ since it is not a solution of $W_+'=0$, by hypothesis) but this is excluded by assumption (i), so $W_-$ is stationary at a point that is not a stationary point of $V$. As we have seen, this implies that $W_-$ is a double-valued function that is not defined for $\sigma>\sigma_-$. The simplest possibility for the second branch of this function is that it continues back to the origin, where it must be identified with $W_M$ for some $M$. 

There is a simple interpretation of this result for the cosmological analog in which the potential is inverted. What was a domain wall becomes a cosmology, represented by the damped motion of a ball in the inverted potential $-V$. The adS maximum at the origin becomes a de-Sitter (dS) minimum and the adS minimum at $\sigma=\sigma_+$ a dS maximum. The condition that the BF bound is satisfied at the adS maximum becomes the condition that motion at the dS minimum is overdamped \cite{Skenderis:2006rr}, so that it approaches monotonically and exponentially fast with exponent $\nu_+$ or $\nu_-$. 
Suppose that the ball is initially at rest at the maximum and (after an infinite time) it falls to the minimum, exponentially fast with exponent $\nu_+$. This cosmological motion corresponds to a domain wall interpolating between the two stationary points of the potential, and it determines a superpotential 
that (a) is single-valued in the interval $[0,\sigma_+]$, (b)  coincides with $W_+$ at the origin, and (c)
is stationary at $\sigma=\sigma_+$. This is the superpotential $W_+$ of the previous paragraph. 
We have assumed its existence but this must require a special choice of $V$ since the ball would typically overshoot, yielding a superpotential with a supersymmetric adS maximum and non-supersymmetric adS minimum.  

Now suppose that we choose initial conditions for a ball on the other side of the dS minimum such that it overshoots  the dS minimum at the origin with non-zero velocity
and continues up the potential towards the maximum. It cannot `just reach' the maximum (because that motion has already been accounted for) so it will either overshoot or it will fall back. There must be some initial conditions for which the latter possibility is realized. As it falls back it will either overshoot again or
it will approach the dS minimum exponentially fast with exponent $\nu_-$. Although the former possibility may be generic, there must exist some choice of initial conditions that realizes the latter. In this case the corresponding domain wall solution determines the double-valued potential for which 
the function $W_-$ defined at the origin is one branch, the other branch being a $W_M$ function for some particular $M$ (which carries the information about the required initial conditions in the mechanical interpretation). The branch point of this double-valued function is the turning point in the motion of the ball as it falls back towards the minimum. 

In the case that the adS maximum is unstable, the cosmological dual situation involves a ball that oscillates indefinitely as it falls to the minimum of the inverted potential. As observed in \cite{Skenderis:2006rr}, this implies an accumulation of zeros of $\dot\sigma$ that prevents a domain wall asymptotic to an unstable adS vacuum from being even piecewise supersymmetric. As we now see, an attempt to construct a superpotential from such a domain wall solution would yield a multi-valued function  with an infinite number of branch points that accumulate at the origin, so there is no superpotential for such a domain wall solution that is differentiable in a neighbourhood of the adS vacuum.

\subsection{E Pluribus Unum?}
\label{sec:unum}

We now turn to multi-scalar models. For a generic multi-scalar model, both the potential and the Killing spinor equation must depend on both the superpotential and the target space metric, both of which will generically depend on all scalar fields,  because this feature is already apparent from the two-scalar models obtained by coupling of a chiral supermultiplet to $d=4$ supergravity. Moreover, there will be additional requirements for supersymmetry arising from the requirement of vanishing supersymmetry variations for the spinor superpartners of the additional fields.  However, a domain-wall solution of a multi-scalar model can always be viewed as a solution of a single-scalar model obtained by truncation after choosing `adapted' coordinates on the scalar field target space for which all scalar fields but one are constant \cite{Celi:2004st}. As the additional conditions required for supersymmetry are just the constancy of the `other' scalar fields, 
the problem of (fake) supersymmetry for multi-scalar model domain-walls has been reduced to the 
single-scalar problem.  

However, there is a difficulty with this  `reduced' notion of fake
supersymmetry (in addition to possible global problems arising from the local nature of the `adapted' target space coordinates).  It may happen, if the single-scalar truncation is inconsistent, that
there are no nearby solutions of the single-scalar model that are also solutions of the multi-scalar model, in which case the fake supersymmetry of the domain wall as a solution of the single-scalar
model is irrelevant to its stability  as a solution of the multi-scalar model. We shall show that  precisely this scenario is realized by the exact solution of subsection \ref{sec:exact}.  

The relation (\ref{sigchi}) satisfied by the axion-domain wall solution of  subsection \ref{sec:exact}
suggests that we choose adapted coordinates $(\tilde\sigma,\tilde\chi)$ 
defined by
\be
e^{\mu\sigma} = \exp\left(\frac{\mu}{\sqrt{J_0}}\tilde\sigma\right) \, J \, \qquad 
\chi -\chi_0  =  \exp\left(-{\mu\over2 \sqrt{J_0}}\tilde\sigma\right) (\tilde\chi + \kappa)\, J^{-1/2}\, , 
\ee
where $\kappa$ is the constant defined in (\ref{Akappa}), and
\be
J  = 1 + \frac{1}{4}\mu^2\left(\tilde\chi + \kappa\right)^2\, , \qquad 
 J_0 = 1+ \frac{1}{4}\mu^2\kappa^2 \equiv \frac{\lambda\left(\lambda-\mu\right)}{\lambda_h^2}\, . 
\ee
Observe that $J_0\ge1$ for $\mu\le\mu_c$, and that the Jacobian of this change of variables is non-zero for all finite $\sigma$, so the adapted coordinates are valid globally as long as there are no identifications on the hyperbolic target space (e.g. those arising from requiring periodicity in $\chi$). 
 In the new coordinates the solution is
\be
\tilde \chi=0\, ,\qquad \tilde\sigma = \sqrt{J_0}\left[\sigma(z) - \frac{1}{\mu} \ln J_0\right]\, ,  \qquad
\varphi = \varphi(z)\, , 
\ee
with $\sigma(z)$ and $\varphi(z)$ as in (\ref{fixedinterior}).  This confirms that the new coordinates are indeed `adapted'  to the solution in the sense of \cite{Celi:2004st}. 

In the gauge used to find the above solution, the effective Lagrangian in the new variables is
\be
2L_{eff} =  e^{\alpha\varphi-\frac{1}{2}\tilde\lambda\tilde\sigma} J^{-\frac{\lambda}{2\mu}}\left( \dot\varphi^2 -\left(J/J_0\right)\dot{\tilde\sigma}^2 -J^{-1}\dot{\tilde\chi}^2 -2\Lambda\right)
\ee
where
\be\label{tildelam}
\tilde \lambda = \lambda/\sqrt{J_0} \equiv \lambda_h \sqrt{\frac{\lambda}{\lambda-\mu}}\, , 
\ee
and the constraint is
\be\label{contilde}
\dot\varphi^2 -\left(J/J_0\right)\dot{\tilde\sigma}^2 -J^{-1}\dot{\tilde\chi}^2 +2\Lambda=0\, . 
\ee
Now observe that
\be
\tilde L_{eff} \equiv \left(J_0\right)^{\frac{\lambda}{2\mu}} L_{eff} \bigg|_{\tilde\chi\equiv 0} =
\tfrac{1}{2} {\tilde f}^{-1} \left(\dot\varphi^2 - \dot{\tilde\sigma}^2\right) - \tilde f \Lambda e^{-\tilde\lambda \tilde\sigma}\, , 
\ee 
where
\be
\tilde f = e^{-\alpha\varphi + \frac{1}{2}\tilde\lambda \tilde\sigma}\, , 
\ee
and that the equation of motion for $\tilde f$, for $\tilde f$ as given, is
\be
\dot\varphi^2 -\dot{\tilde\sigma}^2 +2\Lambda=0\, ,
\ee
which is the constraint (\ref{contilde}) for $\tilde\chi\equiv0$. We see that the `adapted' truncation, achieved by setting $\tilde\chi\equiv 0$, yields a one scalar model with exponential potential. 

The above construction  guarantees that the solution used in the construction of the `adapted'  single-scalar model is a solution of the equations of motion of this model. Let us verify this:  the solution for $(\varphi,\tilde\sigma)$, expressed in terms of $\tilde\lambda$, is such 
that\footnote{From  (\ref{tildelam}) it follows that $\lambda\ne\lambda_h$ as long 
as $\mu\ne0$, as assumed.}
\be
\dot\varphi = \pm \lambda_h \sqrt{\frac{2\Lambda}{\tilde\lambda^2- \lambda_h^2}}\, ,\qquad 
\dot{\tilde\sigma} = \pm \tilde\lambda \sqrt{\frac{2\Lambda}{\tilde\lambda^2- \lambda_h^2}}\, . 
\ee
This is precisely the ``type 1'' fixed point solution of the single-scalar model described in \cite{Sonner:2005sj}, which was shown in that reference to correspond to a flat domain wall solution that is 
supersymmetric with respect to the real superpotential 
\be\label{tildesuperpot}
\tilde W(\tilde\sigma)= \left(\frac{2\Lambda}{\tilde\lambda^2- \lambda_h^2}\right)^{\frac{1}{2}}\exp\left(-\frac{1}{2}\tilde\lambda \tilde\sigma\right)\, . 
\ee

It follows that there is at least one solution of the single-scalar model that is also a solution of the original axion-dilaton model: namely, the solution of the latter used to construct the former. However, it is not guaranteed that any other solution of the single-scalar model will share this property; it will if the truncation is consistent, in the technical sense, but not necessarily if the truncation is not consistent. For the case in hand, the truncation is {\it not} consistent: the expansion of $L_{eff}$ in powers of $\tilde\chi$ shows that there is a linear term, which is such that the field equation for $\tilde\chi$ is satisfied by  $\tilde\chi\equiv0$ iff
\be
\kappa\left(\dot{\tilde\sigma}^2 - 2\Lambda\lambda/\mu\right) =0\, . 
\ee
This is satisfied if $\kappa=0$, but this corresponds to the limiting case in which $\dot\chi=0$, so the truncation is equivalent to setting the axion field to zero in the original Lagrangian; this is evidently a consistent truncation, and we implicitly used this fact in section \ref{sec:constantchi}. Otherwise, the truncation is 
inconsistent because it implies a condition on $\tilde\sigma$. Using the relation between $\tilde\sigma$ and $\sigma$ at $\tilde\chi=0$, and the constraint, we see that this condition states that
\be
\dot\sigma^2 = \lambda_h^2 A^2 \, ,\qquad \dot\varphi^2 = \left(\lambda-\mu\right)^2 A^2\, ,
\ee
where $A$ is the constant defined in (\ref{Akappa}).  As one sees from (\ref{fixedinterior}), this is satisfied by the solution we started with (as was guaranteed) but {\it only} by this solution\footnote{The consistency constraint allows for the sign of $\dot\sigma/\dot\varphi$ to differ from (\ref{fixedinterior}) but there is no solution corresponding to this possibility.}.

\subsection{Axion-dilaton fake supergravity}

The above result suggests that we should extend the notion of fake supersymmetry to models involving at least at axion as well as a dilaton. Here we  consider such an extension, using the coupling of $d=4$ supergravity to a complex chiral superfield as a model.  Consider a Lagrangian density of the form
\be
{\cal L} =\sqrt{-\det g}\left[R -2G\, \partial \tau \cdot \partial\bar\tau -  V\right]\, , 
\ee
where $\tau$ is a complex scalar field taking values in a K\"ahler target space and
\be
G= \partial_\tau \partial_{\bar\tau} {\cal K}(\tau,\bar\tau)\,
\ee
is the K\"ahler target space metric, expressed in terms of the K\"ahler potential ${\cal K}$. We take the scalar field potential $V(\tau,\bar\tau)$ to be given in terms of $K$ and a holomorphic superpotential $P(\tau)$ according to the formula 
\be\label{veeAD}
V= \frac{1}{2}e^{\cal K} \left[ \left|\partial_\tau P + \partial_\tau {\cal K}\,
  P\right|^2 G^{-1}
 - 4\alpha^2\left|P\right|^2\right] \, . 
\ee
This generalizes to arbitrary $d$ the standard $d=4$ formula\footnote{See, e.g.,  \cite{Bagger:1984ge}, but note that this reference uses conventions that differ from ours.}. The potential is invariant under the K\"ahler gauge transformation
\be
K \to K -\left(f+\bar f\right)\, , \qquad P \to e^f P\, , \qquad \bar P \to e^{\bar f} \bar P\, , 
\ee
where $f(\tau)$ is an arbitrary holomorphic function of $\tau$ and $\bar f$ is its anti-holomorphic complex conjugate.  If $P=0$ then $V=0$, so only the $P\ne0$ case is of interest here, but for $P\ne0$ we may choose $P=1$ without loss of generality as this is a K\"ahler gauge choice. 
In this case we have
\be\label{preV}
V= 2 \left[\left|\partial_\tau W\right|^2 G^{-1} - \alpha^2 W^2\right]\, , 
\ee
where
\be\label{WK}
W= e^{K/2}\, . 
\ee
To see that the $d$-dependence is correct we write
\be
\tau = \chi + i\Sigma(\sigma)\, , 
\ee
for some function $\Sigma$ of the dilaton field $\sigma$, and we assume that $K$, and hence $W$, is a function only of $\sigma$, not of $\chi$. In order to get a standard kinetic term for $\sigma$ we must choose the functions $\Sigma(\sigma)$ and $W(\sigma)$ to be related by 
\be\label{WSig}
\left(\frac{W'}{W}\right)' - \left(\frac{\Sigma'{}'}{\Sigma'}\right)
\left(\frac{W'}{W}\right) = \frac{1}{2}\, , 
\ee
where a prime indicates a derivative with respect to $\sigma$. 
Specifically, this equation implies that
\be\label{kinetic}
2G\,  \partial\tau \cdot \partial\bar\tau = \frac{1}{2}\left[
\left(\partial\sigma\right)^2 +  \left(\Sigma'\right)^{-2} 
\left(\partial\chi\right)^2 \right]\, .
\ee
The formula (\ref{preV}) now reduces to the standard $d$-dimensional fake-supergravity formula
(\ref{VW}).

For $d= 4$ mod $4$, the  Killing spinor equation may be generalized to
\be\label{Killeq}
\left[D_\mu -\frac{1}{2}A_\mu \gamma_*- \frac{1}{2(d-2)} e^{K/2}\, \left[ {\rm Re}\,  
P + \gamma_*\, {\rm Im}\,  P\right] \, \Gamma_\mu \right]\epsilon =0\, , 
\ee
where $\gamma_*$ is the product of the $d$ Dirac matrices, satisfying $\gamma_*^2=-1$, and 
\be
A_\mu = -\frac{i}{2}\left(\partial_\mu \tau\partial_\tau K - \partial_\mu\bar\tau \partial_{\bar\tau} K\right)
\ee
is the K\"ahler gauge connection; it transforms as $A \to A -\frac{i}{2}d\left(f-\bar f\right)$. It ensures invariance under K\"ahler gauge transformations because the spinor parameter $\epsilon$ has the  transformation
\be
\epsilon \to \exp\left(-\frac{i}{4}\left(f-\bar f\right)\gamma_*\right)\epsilon\, . 
\ee
When $W$ is a function only of $\sigma$, we have
\be
A_\mu = -\left(\frac{W'}{W\Sigma'}\right) \partial_\mu\chi\, , 
\ee
so it is zero for constant axion. The field strength is 
\be
\partial_{[\mu}A_{\nu]} = - \frac{1}{2\Sigma'} \partial_{[\mu}\sigma \partial_{\nu]}\chi\, , 
\ee
but this vanishes for all domain-wall solutions, irrespective of whether the axion field is constant. 
From this we may conclude that the K\"ahler potential is irrelevant to the integrability conditions required for the existence of a Killing spinor.  As these integrability conditions are all that we will need to consider here, we may ignore the K\"ahler gauge connection in what follows. Setting $P=1$ too, we see that the `effective' Killing spinor equation is precisely (\ref{Kill2}). This verifies that we have the correct $d$ dependence in (\ref{Killeq}), at least for $d=4$ mod $4$; we will not consider here whether there is an appropriate generalization of (\ref{Killeq}) to other dimensions.  

As the potential given by (\ref{VW}) for a superpotential $W$ that is a function only of $\sigma$ is itself only a function of $\sigma$, it is manifest that the $\chi=0$ truncation is consistent, and one then recovers the fake supergravity formalism summarised  at the start of this section. This shows how
the fake supergravity formalism can be recovered from N=1 $d=4$ supergravity, and it checks the
$d$-dependence of our proposed extension of the potential and Killing spinor equations.
However, for what follows it is important  to appreciate that the formula for the potential (\ref{VW}) 
and the Killing spinor equation (\ref{Kill2}) are valid (at least for $d=4$ mod $4$) {\it even if no truncation is made}. All that has been assumed to derive these equations is that $W$ depends only on 
$\sigma$, although the derivation requires that the coupling of $\sigma$ to $\chi$ then be chosen such that  (\ref{WSig}) is satisfied. If, as here, we wish to specify a particular coupling (corresponding to a particular choice of target space) then only certain superpotentials $W$ will be possible.  Specifically, our axion-dilaton model is  found by choosing
\be
\Sigma(\sigma) = \pm \frac{2}{\mu} e^{-\frac{\mu}{2}\, \sigma}\, , 
\ee
in which case (\ref{WSig}) reduces to 
\be\label{WWprime}
\left(\frac{W'}{W}\right)' + \frac{\mu}{2}
\left(\frac{W'}{W}\right) = \frac{1}{2}\, , 
\ee
This equation is solved by\footnote{Other solutions of  (\ref{WWprime}) yield a `non-exponential' potential that is of no particular interest here.}  
\be\label{Wsoln}
W= W_0\, e^{\sigma/\mu}\, , 
\ee
for constant $W_0$, which we assume to be non-zero.  We then find the
potential
\be
V= \Lambda\, e^{-\lambda\sigma}
\ee
where 
\be
\lambda  =-2/\mu \, , \qquad 2\Lambda = 
W_0^2 \left(\lambda^2 - \lambda_h^2\right)\, . 
\ee
What we have now shown is that  a one-parameter  subfamily of  our two-parameter family of axion-dilaton models  may be  considered as `fake axion-dilaton supergravities', in the sense described 
above. This family is defined by $\mu\lambda=-2$ and for $d=4$ there is an underlying `genuine' supergravity theory, which includes the FS model as the special case for which $\lambda=1$.

Actually, we have two separate one-parameter families: one with $\Lambda>0$ and $\lambda>\lambda_h$ and another with $\Lambda<0$ and $\lambda<\lambda_h$. In either case we have $\mu\left(\lambda-\mu\right)<0$ (since $\mu=-2/\lambda<0$) and hence the restriction to the 
one-parameter family is compatible with the existence of the fixed point solution discussed above only if $\Lambda<0$, and it exists in this case only if $\mu\le \mu_c$. In the special case that $\mu=\mu_c$, which is realized by the FS model, we see from (\ref{sigchi}) that $\dot\chi=0$, and hence that the $\chi=0$ truncation $\chi=0$ is consistent. We shall exclude this case as trivial (the solution coincides with one of the fixed-point solutions of section \ref{sec:constantchi}) so we consider only $\mu<\mu_c$, which is equivalent to $\lambda > \lambda_c$ when $\mu\lambda=-2$.
To summarise, a fixed point solution with non-constant $\chi$ exists for $\mu=-2\lambda$ and $\Lambda<0$ provided that 
\be\label{range}
\lambda_c < \lambda < \lambda_h\, . 
\ee
Observe that
\be
W= \sqrt{\frac{2\Lambda}{\lambda^2-\lambda_h^2}} e^{-\frac{1}{2}\lambda\sigma}\, 
\ee
for this 1-parameter family, and that $W$ and $\tilde W$, of (\ref{tildesuperpot}), are identical 
functions of their arguments, as expected since the `adapted' truncation leading to the superpotential $\tilde W(\tilde\sigma)$ coincides with the consistent truncation $\dot\chi\equiv0$ leading to the superpotential $W(\sigma)$ in the limit that $\lambda=\lambda_c$. 

These fixed point solutions are fake supersymmetric in the sense of \cite{Skenderis:2006jq}, which relies on the truncation in `adapted' coordinates of \cite{Celi:2004st}, but are they fake supersymmetric solutions of the untruncated model? We are now in a position to answer this question.
First, we recall from \cite{Sonner:2005sj} that  the integrability
conditions for the Killing spinor equation 
(\ref{Kill2}) are
\be
\dot\varphi = \pm 2\alpha e^{\frac{1}{2}\lambda\sigma} W\, ,\qquad
\dot \sigma = \mp 2 e^{\frac{1}{2}\lambda\sigma} W'\, , 
\ee
and hence, for the superpotential (\ref{Wsoln}), 
\be
\dot\varphi = \pm 2\alpha W_0\, ,\qquad \dot\sigma  = \pm \lambda
W_0\, . 
\ee
Comparing with (\ref{fixedinterior}) we see that supersymmetry
requires $W_0= \sqrt{|\Lambda|}$ and $\mu=\mu_c$ or, equivalently, 
$\lambda=\lambda_c$, which is outside the range  (\ref{range}). 
The fake supersymmetry of the limiting case for which
$\lambda=\lambda_c$ is not surprising because $\dot\chi=0$ in this
case. In all cases for which the axion is not constant, the
single-scalar truncation is inconsistent, the domain-wall solution 
is {\it not} a fake supersymmetric solution of the untruncated model, 
at least not in the sense described here.

%%%%%%%%%%%%%%%%%%%%%%%%%%%%%%%%%%%%
\section{Discussion} \setcounter{equation}{0}
%%%%%%%%%%%%%%%%%%%%%%%%%%%%%%%%%%%%

We have studied domain wall solutions of  gravity coupled to an axion and dilaton, with a hyperbolic target space and exponential dilaton potential. In particular, we have studied the dependence on the 
two parameters $\mu\ne0$ and $\lambda \ge0$ that determine, respectively, the target space radius 
and the dilaton self-coupling. Starting from the observation that domain wall solutions with either (i) constant dilaton or (ii) Minkowski `worldvolume' are essentially determined by the trajectories of an autonomous 2-dimensional dynamical system, we have used the methods of dynamical systems to provide an analysis of the trajectories in these two cases, allowing for either sign of the potential. In case (i) our  results constitute a global completion of previous  results. In case (ii) we have both extended previous results to either sign potential, and provided a global completion of them. In both cases this has been achieved by finding appropriate new variables that allow the global picture to emerge naturally. We have also investigated the nature of the bifurcations that occur when fixed points coincide as the parameters vary. All simple bifurcations are transcritical. We should stress here that these dynamical systems results are equally  applicable to domain-walls and cosmologies, but we have focused here
on the domain-wall interpretation to avoid any confusion that may arise from the flip of the sign of the potential, and curvature $k$, needed to pass from one interpretation to the other. 

One new result of this paper is an exact `scaling' solution for a flat domain wall with non-constant axion field. For one sign of the potential (negative for domain walls but positive for cosmology) this is the domain wall solution that corresponds to the cosmological scaling solution found in  \cite{Sonner:2006yn}. In this context, it was noticed by Rosseel et al. \cite{Rosseel:2006fs} that this solution is unusual in that the motion in the target space is not geodesic. The same observation obviously applies in the domain wall case.

As shown in \cite{Freedman:2003ax,Sonner:2005sj,Skenderis:2006jq}, {\it all} flat or adS-sliced domain wall solutions for which the scalar field is a strictly monotonic function are supersymmetric with respect to a superpotential that is determined by the solution itself.  The monotonicity condition is violated `maximally'  by adS vacua and `badly' by domain walls that are asymptotic to unstable adS vacua. Here we have investigated what happens when it is violated at isolated points, which is a fairly generic  phenomenon. An explicit example confirmed the suggestion of \cite{Skenderis:2006jq} that such domain walls should be `piecewise supersymmetric' with respect to a multi-valued superpotential. It also showed how this is associated with stationary points of the superpotential that are not stationary points of the potential. A nice picture emerged from our further investigation of multi-valued superpotentials in which the domain wall solutions of any given potential determine the global branched structure of the associated superpotential, or superpotentials, and we used this to recover and extend results 
of \cite{Amsel:2007im} on  negative potentials with both a fake supersymmetric maximum and a fake supersymmetric minimum. 

There is a sense in which the results on supersymmetry of domain walls in single-scalar models carry over, at least locally, to multi-scalar models because any domain-wall solution of a multi-scalar model is also a solution of the single-scalar model obtained by truncation in `adapted'  target-space coordinates \cite{Celi:2004st}. Obviously, one cannot expect to use fake supersymmetry in this sense to prove stability with respect to fluctuations of the truncated scalars but one might hope that it would be sufficient for stability against fluctuations of the one untruncated scalar. The results of this paper show that this hope is unfounded, in general. The problem  is that the `adapted' truncation is not guaranteed to be a consistent one, and if it is not consistent then the `nearby' (and generically, time-dependent) solutions of the single-scalar model (needed to establish stability in the single-scalar context) do not lift to `nearby' solutions of the original multi-scalar model. It was shown in \cite{Celi:2004st}, in the context of $d=5$ supergravity,  that this problem does not arise if the original multi-scalar solution is supersymmetric, so only a non-supersymmetric domain-wall solutions of $d=5$ supergravity could lead to an inconsistent adapted truncation. It would be a surprise if this result were not generally valid, and we conjecture that the converse is true too: that inconsistency of an adapted truncation is associated with a failure of the domain wall to be supersymmetric as a solution of an untruncated supergravity theory. 

We tested this conjecture on  the new scaling solution mentioned above. This solution has the interesting feature, which we suspect is related to its non-geodesic character, of leading to an 
{\it inconsistent} one-scalar truncation.  The test was made possible by a derivation of the $d=4$ fake supergravity formalism from minimal $d=4$ supergravity coupled to a chiral supermultiplet,  and an extension of this formalism to cover a one-parameter subfamily of our original two-parameter 
family of axion-dilaton models, which could therefore be viewed as fake axion-dilaton supergravity theories. The scaling solution was found to be non-supersymetric in this context, as the conjecture requires.

%%%%%%%%%%%%%%%%%%%%%%%%%%%%%%%%%%%%
\section*{Acknowledgements}
%%%%%%%%%%%%%%%%%%%%%%%%%%%%%%%%%%%%

We thank Alessio Celi, Jonathan Dawes, Joaquim Gomis, Kostas Skenderis and Antoine Van Proeyen for helpful conversations. 

\setcounter{section}{0}
%%%%%%%%%%%%%%%%%%%%%%%%%%%%%%%%%%%%%%%%%%%%%%%%%%%%%%%%%%
\bigskip

%%%%%%%%%%%%%%%%%%%%%%%%%%%%%%%%%%%%
%%%%%%%%%%%%%%%%%%%%%%%%%%%%%%%%%%%%

%\bibliography{geo}
%\bibliographystyle{abbrv}
\end{document}